%% file: main.tex
\documentclass[aps,prb,twocolumn,groupedaddress,10pt,longbibliography]{revtex4-2}

\usepackage{natbib}
\usepackage{graphicx}
\usepackage{xcolor}
\usepackage{siunitx}
\usepackage{amsmath}
\usepackage{amsmath}
\usepackage{amssymb}
\usepackage[ruled,lined]{algorithm2e}
\usepackage{hyperref}
\usepackage{fancyhdr}
\usepackage{silence}
\DeclareSIUnit\angstrom{\text {Å}}

\begin{document}

\title{Magnetothermal Properties with Sampled Effective Local Field Estimation}

\author{Nicholas Brawand}
\email{nbrawand@mitre.org}

\author{Nima Leclerc} 

\author{Emiko Zumbro}
\email{ezumbro@mitre.org}
\affiliation{The MITRE Corporation, 7525 Colshire Dr, McLean, VA, 22102, USA}

\date{\today}

 \begin{abstract}
We introduce a first-principles method for predicting the magnetothermal properties of solid-state materials, which we call \emph{Sampled Effective Local Field Estimation}. This approach achieves over two orders of magnitude improvement in sample efficiency compared to current state-of-the-art methods, as demonstrated on representative material systems. We validate our predictions against experimental data for well-characterized magnetic materials, showing excellent agreement. The method is fully automated and requires minimal computational resources, making it well suited for integration into high-throughput materials discovery workflows. Our method offers a scalable and accurate predictive framework that can accelerate the design of next-generation materials for magnetic refrigeration, cryogenic cooling, and magnetic memory technologies.
\end{abstract}

\maketitle
\thispagestyle{fancy}
\section{Introduction} \label{Introduction}
Magnetic materials are vital to many modern technologies, including cryogenic cooling systems \cite{Gschneidner2003, Numazawa2002}, memory and logic devices \cite{manipatruni2019scalable}, and magnetic-field delivery systems. Each of these applications rely on magnetothermal properties of the active materials, such as the magnetocaloric effect or magnetic heat capacity at specific operating temperatures. For instance, cryogenic cooling systems used in infrared imaging \cite{han2022cryogenic}, quantum information science, and quantum sensors \cite{Johnson2022} require antiferromagnetic materials like HoCu$_2$ that exhibit exceptionally high magnetic specific heat below 10 K \cite{Gschneidner2003}. Addressing this need calls for reliable and efficient tools to predict magnetothermal properties, enabling the discovery and design of materials tailored to precise operational requirements.

Current state-of-the-art computational methods that predict temperature dependent effects, such as magnetic specific heat \cite{Rajan1983,Ghivelder1999}, magnetic susceptibility \cite{Ohno1997,Mauri1996}, magnetocrystalline anisotropy \cite{Daalderop1990}, and the magnetocaloric effect \cite{Cooley2023} scale poorly in sample and time efficiency, and as such, are not reliable in high-throughput settings. These methods often demand sampling over combinatorially large spaces of magnetic microstates \cite{Shang2010} to accurately model the statistical distribution of magnetic configurations at desired temperatures. Other proposed first-principles methods rely on expensive total-energy calculations of magnetically constrained supercells computed with spin-polarized density functional theory (DFT)  \cite{Kvashnin2015}. The limited accuracy, reliance on user-defined parameters, and poor computational scaling of existing predictive approaches have hindered their integration into automated materials design workflows~\cite{jain2020materials}.   

Magnetic materials exhibit lattice, magnetic, and magnetostructural degrees of freedom, each contributing to the total specific heat, expressed as \( c_{\text{total}} = c_l + c_m + c_{ms} \). While the lattice contribution \( c_l \) can be reliably computed from the phonon density of states using DFT~\cite{togo2023first, sondenaa2007heat}, robust first-principles approaches for estimating the magnetic \( c_m \) and magnetostructural \( c_{ms} \) contributions remain lacking. Previous methods have made progress toward this end but with limitations \cite{lin2000magnetic, tishin1999magnetocaloric, Walsh2022}. Some approaches compute the magnetic Gibbs free energy relying on sampling over hundreds of spin-constrained DFT calculations \cite{Mendive-Tapia2022}, while others rely on semi-empirical fits to DFT data using phenomenological models \cite{Sangeetha2016}, which can lack generalizability as they depend on user-defined parameters. Monte Carlo-based frameworks, such as those developed by Walsh \emph{et al}. \cite{Walsh2022}, have shown strong performance but can be computationally intensive. 

In this Letter, we introduce Sampled Effective Local Field Estimation (SELFE), a method for predicting magnetothermal properties from first principles. SELFE combines DFT-derived exchange interactions with a statistical framework inspired by prior single-site spin sampling models \cite{Mendive-Tapia2022}, extending them to capture two-body interactions without requiring spin-constrained supercell calculations. This allows SELFE to model the competition between enthalpic and entropic contributions to the free energy of a magnetic material. Our approach employs Green’s function techniques to compute exchange parameters from first principles for a given material, avoiding the need for free parameters. This makes SELFE readily automatable and well suited to integrate into large-scale materials discovery pipelines \cite{horton2019high}.

This Letter is organized as follows. First, we provide a theoretical grounding to our approach, introducing the SELFE algorithm, and explain its differentiation with respect to existing state-of-the-art methods.  Second, we consider the physical systems: face centered cubic (FCC)-Fe and body centered cubic (BCC)-Fe. We present empirical results of SELFE applied to each, and compare against experimental baselines and other computational methods.  Lastly, we discuss the behavior and performance of our approach and our outlook to integrate this method into a broader materials discovery framework. 

\section{Theory} \label{Theory}

Conventionally, the magnetothermal properties of magnetic materials have been studied with a Hamiltonian describing the collection of single-site internal magnetic fields shown in Eq.(\ref{eq:trial_hamiltonian}), initially introduced by Gyorffy \emph{et al.} \cite{gyorffy1985first}.

\begin{equation}\label{eq:trial_hamiltonian}
\mathcal{H} = -\sum_{i=1}^N \mathbf{h}_i  \cdot \hat{\mathbf{e}}_i  
\end{equation}

\noindent
where $N$ is the number of magnetic sites in the lattice, $i$ is the site index, $\hat{\mathbf{e}}_i$ is the normalized three-dimensional magnetic moment unit vector (parameterized by polar and azimuthal angles $(\theta_i,\phi_i)$), and $\mathbf{h}_i$ is the onsite interaction term. Because of the single-site nature of the above Hamiltonian, the partition function can be evaluated analytically (see Ref. \cite{Mendive-Tapia2022} and the Supplemental Material of this work) and the conditional probability distribution of the magnetic moment at each magnetic site is given by:

\begin{equation}\label{eq:pdf}
   p(\theta_i|\beta, h_i) = \frac{\beta h_i}{4\pi \sinh{(\beta h_i)}} e^{\beta h_i \cos{(\theta_i)}}
\end{equation}

\noindent
where $\theta_i$ is the polar angle of the magnetic moment at the $i$th site relative to $\mathbf{h}_i$. $p(\theta_i|\beta,h_i)$ is also impacted by the local field strength $h_i=|\mathbf{h}_i|$ and $\beta = \frac{1}{k_b T}$, where $T$ is the temperature and $k_b$ is Boltzmann's constant. The azimuthal angle at each site $\phi_i$ is then treated independently from $\theta_i$ and generated relative to $\mathbf{h}_i$. $\phi_i$ is sampled from $U(0, 2\pi)$ where $U$ denotes the uniform distribution with lower and upper bounds of $0$ and $2\pi$.

At a given temperature $T$ and for the set of material-dependent site-level effective fields $\{h_i\}$, the configuration of magnetic moments $\{\hat{\mathbf{e}}_i\}$ can be generated by sampling $\{(\theta_i, \phi_i)\}$ from the probability distributions $p(\theta_i|\beta, h_i)$ and $U(0, 2\pi)$. Sampling the set of site-level angles $\theta_i \sim p_i(\theta_i|\beta, h_i)$ and $\phi_i \sim U(0, 2\pi)$ in this manner allows us to generate magnetic configurations in a physically meaningful way, where the probability distribution for sampling a single magnetic configuration $\{(\theta_i, \phi_i)\}$  at temperature $T$ can be written as $p (\{(\theta_i, \phi_i)\} | \beta , \{ h_i\} )  = \prod_{i=1}^N p(\theta_i|\beta, h_i)$ with $N$ as the total number of magnetic sites.  

Previous implementations have aimed to estimate the partition function directly by generating sets of angular coordinates $\{(\theta_i, \phi_i)\}$ with this procedure to perform computationally costly DFT calculations of magnetically constrained supercells with 100s of atoms \cite{Mendive-Tapia2022, Shang2010}. These approaches have also required the user to specify the site-level fields $\{h_i\}$  empirically as free parameters, while their true values depend on local exchange interactions that should be treated quantum mechanically for unknown materials \cite{anderson1950antiferromagnetism, lines1964green}. SELFE overcomes both of these limitations by calculating the effective fields $\textbf{h}_i$ purely from first principles and then uses these to parameterize a self-consistent statistical model to recover the desired magnetothermal properties.   

Rather than treating site-level mean field $h_i$ as free parameters, SELFE estimates the set of $\{ h_i\}$ values from their atomistic level origins. Using a Green's function-based method \cite{korotin2015calculation} to model an effective two-body interaction for each magnetic site, SELFE circumvents the need to directly estimate the partition function over magnetic configurations. At the microscopic level, phase transitions in magnetically ordered systems arise from exchange interactions between neighboring magnetic moments causing spins to orient in opposite directions in antiferromagnets, to align in the same direction in ferromagnets, and to be randomly oriented in paramagnets. The interplay between material-dependent exchange coupling $J_{ij}$ (for magnetic sites $i$ and $j$) and lattice temperature $T$ produces unique energetically favorable magnetic configurations $\{\hat{\textbf{e}}_i\}$ leading to distinct magnetic phases and enabling the prediction of material properties such as magnetic specific heat.

Unlike previous approaches that require numerous magnetically constrained spin supercell calculations, SELFE requires only a single total energy DFT calculation on the unit cell to obtain the ground-state wave function. These wave functions are then used to obtain orbital-to-orbital interactions and the exchange interaction between lattice sites $i$ and $j$ in the following way. First, we identify site-level states, denoted by $|w_{n\mathbf{k}} \rangle$ where $n$ is the orbital index and $\mathbf{k}$ is the momentum vector. We then follow the procedure introduced by Korotin \emph{et al.} \cite{korotin2015calculation} to compute the isotropic exchange interaction strength $J_{ij}$ using overlap integrals and Green's functions associated with pairs of sites $i$ and $j$ as in Eq. (\ref{eq:green}).

\begin{equation}\label{eq:green}
\begin{split} 
 G_{ij, \sigma}^{mn} (\epsilon) = \int_{\Omega_{BZ}} d\mathbf{k} e^{i\mathbf{k} \cdot (\Delta \mathbf{R}_i - \Delta \mathbf{R}_j ) }   (\epsilon + E_F - H_{mn, ij, \sigma}^{WF} ( \mathbf{k})  )^{-1}    \\ 
 J_{ij}  = -\frac{1}{2\pi} \int_{-\infty}^{E_F} d\epsilon \sum_{m, n, o,p}  \Im (\Delta_{i}^{mn} G_{ij, \downarrow}^{no}\left(\epsilon\right)    \Delta_j^{op} G_{ji, \uparrow}^{pm}\left(\epsilon\right))  
\end{split} 
\end{equation} 

\noindent
where $G_{ij, \sigma}^{mn} (\epsilon)$ denotes the inter-site Green's function evaluated at energy $\epsilon$, and $\sigma$ is the spin quantum number, which describes the interaction between orbitals $m$ and $n$ at lattice sites $i$ and $j$ (e.g. this could describe the interaction between the $d_{x^2 - y^2}$ and $d_{xy}$ orbitals on lattice sites A and B of FCC-Fe). Note that the Green's function is integrated  in $\mathbf{k}$-space over the Briollion Zone (denoted as $BZ$). $E_F$ is the Fermi energy, $\Delta \mathbf{R}_i$ and $\Delta \mathbf{R}_j$ are the differences between the positions of sites $i$ and $j$ in their respective lattices relative to the primitive cell, $H_{mn, ij, \sigma}^{WF} ( \mathbf{k})$ is the single-electron matrix element describing the coupling between orbitals with indices $m$ and $n$ for lattice sites $i$ and $j$. The Green's function $G_{ij, \sigma}^{mn}$ is used to compute the site-to-site exchange interaction $J_{ij}$ as shown in Eq. (\ref{eq:green}), which involves a summation over all orbital states (corresponding to indices $m, n, o, p$ between sites $i$ and $j$. $\Delta_i^{mn}$ ($\Delta_j^{op})$  correspond to the site-level coupling strengths between different orbitals, given by $\Delta_k^{mm^\prime} = \int_{BZ} (H_{ii, mm^\prime, \uparrow}^{WF} (\mathbf{k}) - H_{ii, mm^\prime, \downarrow}^{WF} (\mathbf{k}) ) d\mathbf{k}$. This now provides a procedure to systematically predict the site-level magnetic exchange interaction from first principles, which can be leveraged to parameterize a statistical model to predict magnetothermal properties of interest.  

Using a set of material-dependent $\{J_{ij}\}$ for all pairs of sites, we can rewrite the effective site-level Hamiltonian in Eq. (\ref{eq:trial_hamiltonian}) as a Heisenberg-type formulation \cite{Walsh2022} shown in Eq. (\ref{eqn:two_body_h}), now summing over pairs of lattice sites corresponding to indices $i$ and $j$. This goes beyond existing approaches, replacing the user-defined free parameter $h_i$ with a material-dependent interaction $J_{ij}$ that can be computed from first principles.

\begin{equation}\label{eqn:two_body_h}
\mathcal{H} = -\sum_{i=1}^N 
                  \underbrace{\sum_{j=1}^N J_{ij} \hat{\mathbf{e}}_j}
                            _{\bar{\mathbf{h}}_i(\{\hat{\mathbf{e}}\})} 
                            \cdot \hat{\mathbf{e}}_i   
\end{equation}  

\noindent
where $\bar{\mathbf{h}}_i(\{\hat{\mathbf{e}}\}) = \sum_{j=1}^N J_{ij} \hat{\mathbf{e}}_j$ is the effective local field at the $i$th site. Because the form is the same as the original one body Hamiltonian, we can use the same probability distribution above (Eqn. \ref{eq:pdf}) to sample the magnetic moment angles $\theta_i$ and $\phi_i$. We can sample from these distributions in a self-consistent manner, resampling angles $\theta_i \sim p(\theta_i | \beta, \mathbf{h}_i)$ and $\phi_i \sim U(0,2\pi)$ to compute a new $\bar{\mathbf{h}}_i(\{\hat{\mathbf{e}}_i\})$ and this process is repeated until convergence in the probability distributions is reached. Following convergence, we proceed to compute the magnetothermal properties of interest. In essence, this captures the SELFE method that we introduce in this Letter.      

\begin{figure}[!ht]
\centering
\includegraphics[scale=0.4]{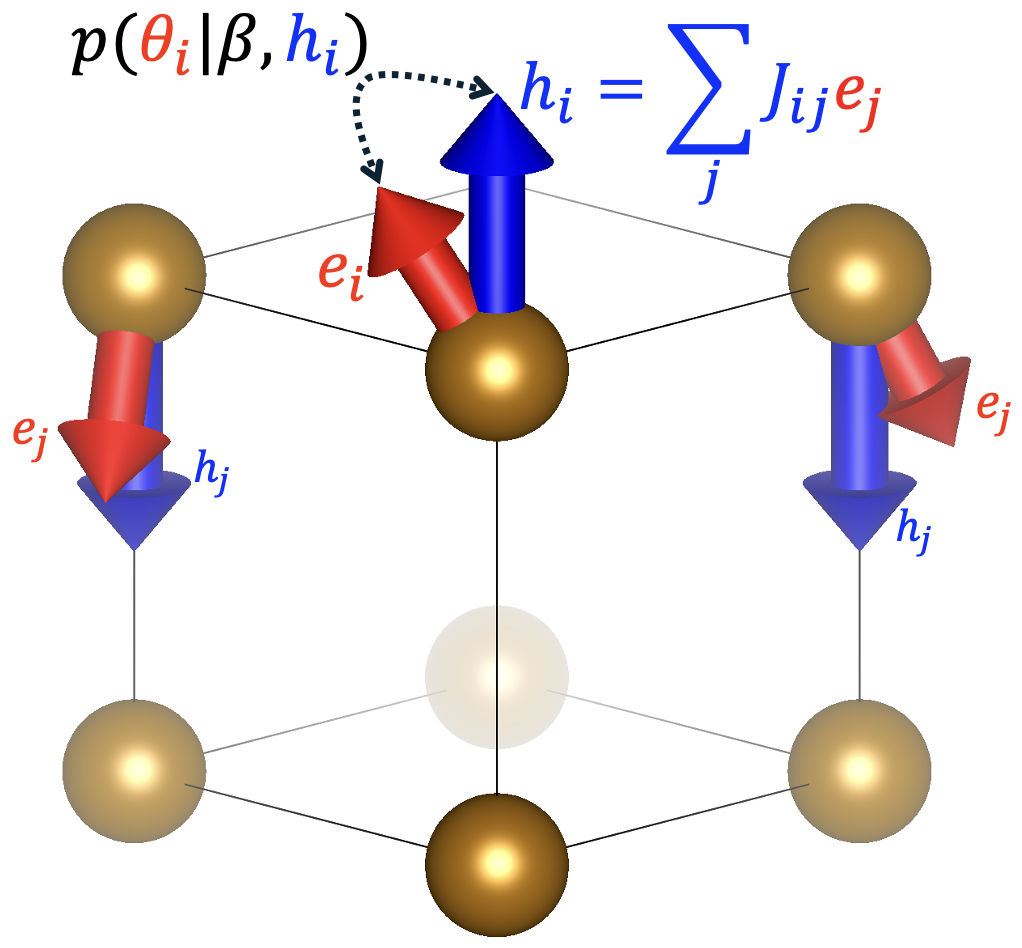}
\caption{Visual representation of Sampled Effective Local Field Estimation concepts. $p(\theta_i|\beta, h_i)$ is the distribution of the polar angles $\theta_i$ of the $i$th site impacted by the effective local field $h_i$.  $h_i$ is determined by other spins $e_j$ and exchange parameters $J_{ij}$. $p(\theta_i|\beta,h_i)$ is also impacted by temperature-dependent $\beta = \frac{1}{k_b T}$.}
\label{fig:toc}
\end{figure}

We qualitatively illustrate the SELFE concept in Fig. (\ref{fig:toc}), which shows the crystal structure of a magnetic material, where each lattice site is given an effective magnetic field $\mathbf{h}_i$ shown in blue and a magnetic moment vector $\mathbf{e}_i$ shown in blue. We see that $\mathbf{h}_i$ is influenced by its interactions with other magnetic sites, mediated through the intrinsic exchange interactions $J_{ij}$. Intuitively, this shows how the effective field (magnitude and direction) is influenced by its neighboring sites at a given temperature and these orientations are determined by the collection of site-level probability distributions $p(\theta_i | \beta, h_i)$ in a self-consistent manner until convergence is reached.     

Running the algorithm self-consistently, we aim to generate a converged set of effective fields $\{ \bar{\mathbf{h}}_i(\{\hat{\mathbf{e}}\})\}$ for all magnetic sites, requiring a metric for convergence of the distribution over these sites. In SELFE, we employ a Z-score based metric commonly used in statistics \cite{mare2017nonstationary, ahsanullah2014normal} as our convergence criterion. Specifically, we monitor the distribution of total energy differences between successive iterations, defined as \(\Delta_i = E_i - E_{i-1}\), and assess whether their mean is statistically indistinguishable from zero. Convergence is declared when the normalized test statistic falls below a threshold determined by the Student's t-distribution, indicating that the energy has stabilized. Once this criterion is met, the resulting magnetic configuration is considered energetically favorable and can be used to compute magnetothermal properties of the material.

Our approach is outlined in Algorithm~\ref{alg:SELFE}. We begin by initializing a random configuration of normalized site-level magnetic moments for \( N \) lattice sites, denoted by \( \{ \hat{\mathbf{e}}_i \} \in \mathbb{R}^{N \times 3} \). We also specify the exchange interaction matrix \( \{ J_{ij} \} \in \mathbb{R}^{N \times N} \), computed using the Green's function method described above, as well as the following parameters: the mixing factor \( \gamma \), the maximum number of iterations \( n_{\max} \), the window size \( n \) used to compute energy difference statistics, and the confidence level \( \alpha \), which sets the threshold for convergence. Once initialized, the algorithm enters a self-consistent loop until the convergence criterion is met.  

For each iteration of SELFE, we parameterize Eq. (\ref{eqn:two_body_h}) with the set of pairwise exchange values $\{J_{ij}\}$, producing an estimate of the set of the calculated effective fields $\{\bar{\mathbf{h}}_i\} \in \mathbb{R}^{N\times 3}$ at each lattice site. 

For a single iteration of the self-consistent algorithm, the fields  $\bar{\mathbf{h}}_i$ are then used to parameterize the distribution over polar angles for site $i$ by $p(\theta_i | \beta, h_i)$ defined in Eq. (\ref{eq:pdf}). Azimuthal angles $\phi_i$ are then sampled over $U(0, 2\pi)$ to generate a pair of spherical coordinates $(\theta_i, \phi_i)$ defining a set of magnetic moment unit vectors $\{ \hat{\mathbf{e}}_i^\prime \}$. 

As shown in Fig. \ref{fig:toc}, $\theta_i$ and $\phi_i$ are generated with respect to the local reference frame, where the local z axis points along the direction of $\mathbf{\bar{h}}_i$ for each lattice site. To express the magnetic moment coordinates in the global reference frame of the crystal, we apply a rotation matrix ($R_h^0$). This matrix transforms vectors from the local $\mathbf{\bar{h}}_i$ frame to the global coordinate frame, ensuring consistency across all lattice sites. Specifically, the transformation is given by $\mathbf{\hat{e}}_i = R_{h}^{0} \cdot \mathbf{\hat{e}}'_i$, where $\mathbf{\hat{e}}'_i$ is in the local reference frame of the $i$th lattice site and $\mathbf{\hat{e}}_i$ is in the global reference frame. Further details about the construction of the rotation matrix can be found in the Supplemental Material.

The transformed moment vector in the global reference frame $\hat{\mathbf{e}}_i^k$ is then mixed with the vector obtained from the previous iteration $\hat{\mathbf{e}}_i^{k-1}$ using a standard mixing procedure $\hat{\mathbf{e}}_i^{k+1} = \gamma \hat{\mathbf{e}}_i^{k} + (1-\gamma )\hat{\mathbf{e}}_i^{k-1}$ with $\gamma \in [0,1]$. Doing this for the collection of magnetic sites $\{\hat{\mathbf{e}}_i\}$, the Curie-Weiss relation Eq. (1) is employed to obtain the total energy $E_k$ at iteration $k$. This procedure is shown in Alg. \ref{alg:SELFE}.

\begin{algorithm}
\caption{Self-Consistent Sampled Effective Local Field Estimation with Z-score Convergence}\label{alg:SELFE}
\KwIn{$\{\hat{\mathbf{e}}_i\}$, $\{J_{ij}\}$,  $\alpha$ , $n$, $n_{\max}$, $\gamma$}
\tcp{Initialize}
$E \gets \text{zeros}(n_{\max})$\;
$Z_0 \gets F_{t, n-1}^{-1} \left(1 - \frac{\alpha}{2}\right)$\;
$Z \gets Z_0$\; 
$k \gets 0$\;
\While{$\left(|Z| \ge Z_0\right) \land \left(k \leq n_{\max} \right)$}{
    \tcp{Update Physical Quantities}
    $ \{ \bar{\mathbf{h}}_i \}_{i=1}^N  \gets \left\{\sum_{j=1}^N J_{ij} \hat{\mathbf{e}}_j\right\}_{i=1}^N $ \;
    $\{ \hat{\mathbf{e}}_i^\prime \}_{i=1}^N \gets
    \left\{ 
    \text{sph}\left(\theta_i, \phi_i\right)
    \;\middle|\;
    \begin{bmatrix}
    \theta_i \sim p(\theta_i \mid \beta, \bar{\mathbf{h}}_i) \\
    \phi_i \sim \mathcal{U}(0, 2\pi)
    \end{bmatrix}
    \right\}_{i=1}^N$\;
    $\{\hat{\mathbf{e}}_i\}_{i=1}^N \gets \left\{\gamma 
     R_{h}^0 \hat{\mathbf{e}}_i^\prime + (1-\gamma )\hat{\mathbf{e}}_i\right\}_{i=1}^N$\;
    $E_k \gets -\sum_{i=1}^N \bar{\mathbf{h}}_i \cdot \hat{\mathbf{e}}_i$\;
    
    \tcp{Update Convergence Criteria}
    \If{$k\geq n$}{
        $\{\Delta_m\}_{m=1}^n \gets \left\{E_k-E_{k-m}\right\}_{m=1}^n$\;
        $\bar{\Delta} \gets \frac{1}{n} \sum_{m=1}^n \Delta_m$\;
        $\sigma_{\Delta}^2 \gets \frac{1}{n} \sum_{m=1}^n (\Delta_m-\bar{\Delta})^2$\;
        $Z \gets \frac{\bar{\Delta}}{\sigma_{\Delta}} \sqrt{n}$\;
    }
    $k = k + 1$\;
}
\tcp{Return Final Orientations \& Energy}
\Return{$\{\hat{\mathbf{e}}_i\}, E_{k{-}1}$}
\end{algorithm} 

We assess the convergence of SELFE by tracking the recent energy differences across iterations. Specifically, we define
\begin{equation}
\Delta_m = E_k - E_{k-m}
\end{equation}

for the last \(n\) iterations, where \(E_k\) is the energy at iteration \(k\). To quantify convergence, we compute the normalized statistic
\begin{equation}
Z = \frac{\bar{\Delta}}{\sigma_{\Delta}} \sqrt{n},
\end{equation}

where \(\bar{\Delta}\) is the mean of the \(\Delta_m\), and \(\sigma_{\Delta}\) is the standard deviation. Under the assumption of approximately independent and normally distributed energy differences, this statistic follows a Student’s t-distribution with \(n - 1\) degrees of freedom.

We compare the magnitude \(|Z|\) to the critical value

\begin{equation}
Z_0 = F^{-1}_{t, n-1}(1 - \alpha / 2),
\end{equation}

and declare convergence when \(|Z| < Z_0\), corresponding to a two-sided confidence level of \(1 - \alpha\), ensuring that deviations in both directions are accounted for. Here, \( F^{-1}_{t, n-1}(\cdot) \) denotes the inverse cumulative distribution function of the Student's t-distribution with \( n - 1 \) degrees of freedom. This approach enables robust and interpretable convergence detection by leveraging statistical principles to quantify fluctuations in energy, offering a principled measure of stability in the iterative process.

To compute thermodynamic properties, we fix a temperature and iteratively obtain self-consistent values of $\bar{\mathbf{h}}_i$ and $\mathbf{\hat{e}}_i$ using Algorithm~\ref{alg:SELFE}. This process is repeated, with each iteration drawing new samples initialized from the previous configuration. Statistical quantities are then estimated from the sampled configurations. For example, the critical temperature is evaluated by computing the variance of the total energy $E$ across samples collected at each temperature. This procedure is demonstrated in the following section.

\section{Computational Details}

DFT calculations were performed to obtain the electronic structure using the primitive cells of all systems to arrive at $\{J_{ij}\}$ using the approach introduced earlier. The resulting $\{J_{ij}\}$ wre incorporated into both the SELFE framework presented above and, for comparison, into the established spin dynamics package Multibinit \cite{martin2018multibinit} as a baseline. 

Spin polarized ground state DFT calculations were performed in Siesta \cite{soler2002siesta}. Input files are provided in the Supplemental Material of this work. All DFT calculations used Optimized Norm-Conserving Vanderbilt (ONCV) pseudopotentials \cite{Garc_a_2018} and the Perdue-Burke-Ernzerhof (PBE) exchange correlation functional \cite{perdew1996generalized}.  $k$-point mesh was set to 14x14x14. Localized Double $\zeta$ Polarized (DZP) basis sets were used with a pseudoatomic orbitals (PAO) energy shift of 0.1 eV to optimize the basis set. A mesh cutoff of 2721.14 eV (200 Ry) was used to define the grid for real-space integrals. Isotropic exchange interaction parameters were calculated with Eq. (\ref{eq:green}) using the TB2J package \cite{he2021tb2j}.           

Provided the set of isotropic exchange parameters $\{J_{ij}\}$, a 5x5x5 spin lattice was constructed for each system and the magnetic contribution to the specific heat $c_m$ was computed  using the Landau-Lifshitz-Gilbert equations \cite{lakshmanan2011fascinating} as implemented in Multibinit. The Multibinit results were obtained after 5,000 (BCC-Fe) and 2,000 (FCC-Fe) thermalization  steps and 50,000 (BCC-Fe) and 10,000 (FCC-Fe) averaging steps. SELFE calculations were performed by sampling 350 spin samples from $p(\theta_i|\beta, \bar{\mathbf{h}}_i)$ while maintaining self-consistency using Alg. \ref{alg:SELFE} with $\alpha = 0.05$ and $n = 5$. This clearly demonstrates the advantages in sample efficiency of SELFE compared to Multibinit.   

\section{Results} \label{Results}
\begin{figure}[!ht]
\centering
\includegraphics[scale=1.0]{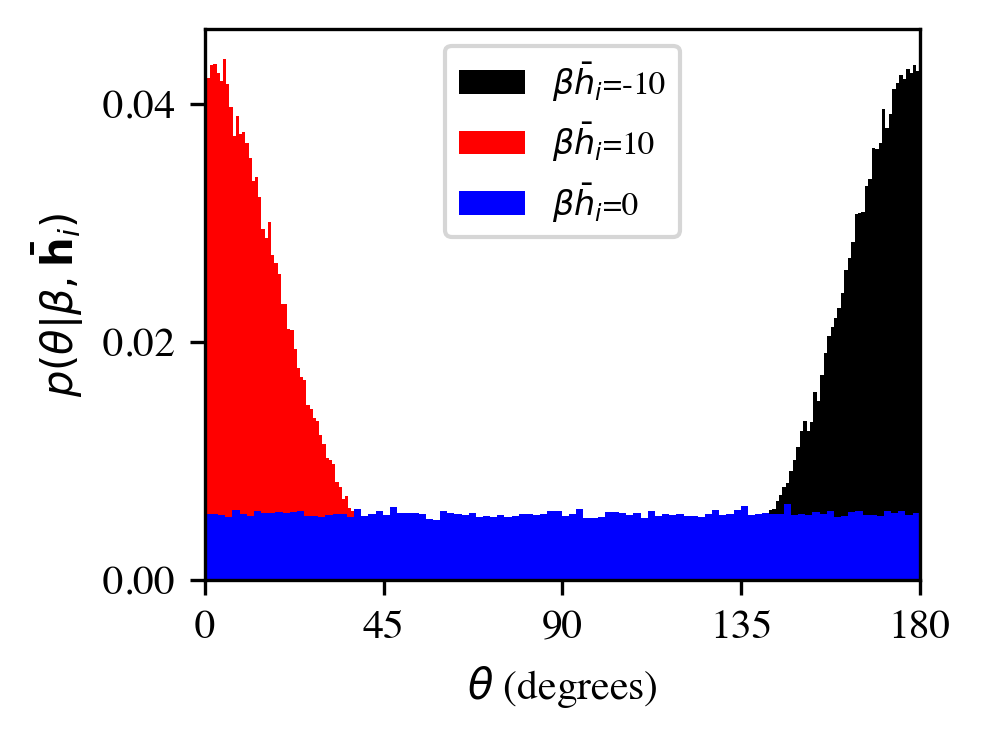}
\caption{Sampling distributions for the polar spin angle $\theta$ for three different absolute effective local field values $\bar{h}_i$ sampled from the fitted inverse cumulative distribution function $\Phi^{-1}(x)$ constructed from Eqn. \ref{eq:pdf}. Positive and negative values of $\bar{h}_i$ yield angles concentrated around $0^\circ$ and $180^\circ$, respectively, indicating that $\theta$ is appropriately influenced by the effective local field. As $\bar{h}_i$ approaches zero, the distribution becomes uniform, reflecting the absence of a strong local field influence.}
\label{fig:dist}
\end{figure}

To sample from the distribution $p(\theta_i|\beta, \bar{\mathbf{h}}_i)$ directly, we fit its inverse cumulative distribution $\Phi^{-1}(x)$ numerically for each magnetic site. The angles $\theta_i$ are then sampled by evaluating $\Phi^{-1}(x)$, where $x$ is drawn from a uniform distribution, $x \sim U(0, 1)$. The sampling distribution of $\theta_i$ is shown in Fig.  \ref{fig:dist} for multiple values of $\beta \bar{h}_i$, where $\bar{h}_i = \text{sign}(\bar{\mathbf{h}}_{i}\cdot \mathbf{\hat{e}}_z)|\bar{\mathbf{h}}_i|$ and $\mathbf{\hat{e}}_z$ is the $z$ direction component of $\mathbf{\hat{e}}_i$.

As expected, positive and negative values of $\bar{h}_i$ corresponding to ferromagnetic (FM) and antiferromagnetic (AFM) behavior, yield angles concentrated around $0^{\circ}$ and $180^{\circ}$, respectively. We also note that the distribution becomes uniform  when $\beta h_i =0$ (e.g., at high temperatures) corresponding to a paramagnetic phase. This indicates that $\theta$ is correctly influenced by the effective local field.  

As a first application of SELFE, we compute the theoretical N\'eel ($T_N$) and Curie ($T_C$) temperature of FCC-Fe and BCC-Fe. Iron is a prototypical magnetic material that exhibits distinct structural and magnetic properties depending on its crystallographic phase. The two most common phases of iron are the body-centered cubic (BCC) and face-centered cubic (FCC) structures, each with a unique magnetic order.  

The BCC phase of iron is well-established as ferromagnetic (FM) at temperatures below its Curie temperature. Experimentally, the Curie temperature of BCC-Fe is approximately 1,043 K, with theoretical predictions yielding similar values, such as 1,060 K \cite{haglof2021calphad}. The FCC phase of iron, while metastable at low temperatures, can be artificially stabilized; finely dispersed FCC-Fe precipitates in Cu exhibit antiferromagnetic (AFM) behavior, with a critical temperature estimated to be around 67 K \cite{he2022third}. Theoretical studies propose two AFM configurations for FCC-Fe: a single-layer AFM structure (AFM-I) and a double-layer AFM structure (AFM-D), with estimated Néel temperatures (\(T_N\)) of 192 K \cite{haglof2021calphad, he2022third, li2021effects}. In this Letter, we leverage the simplicity and contrasting magnetic orderings of BCC-Fe and FCC-Fe as benchmarks for validating the SELFE method.

\begin{figure}[!ht]
\centering
\includegraphics[scale=0.55]{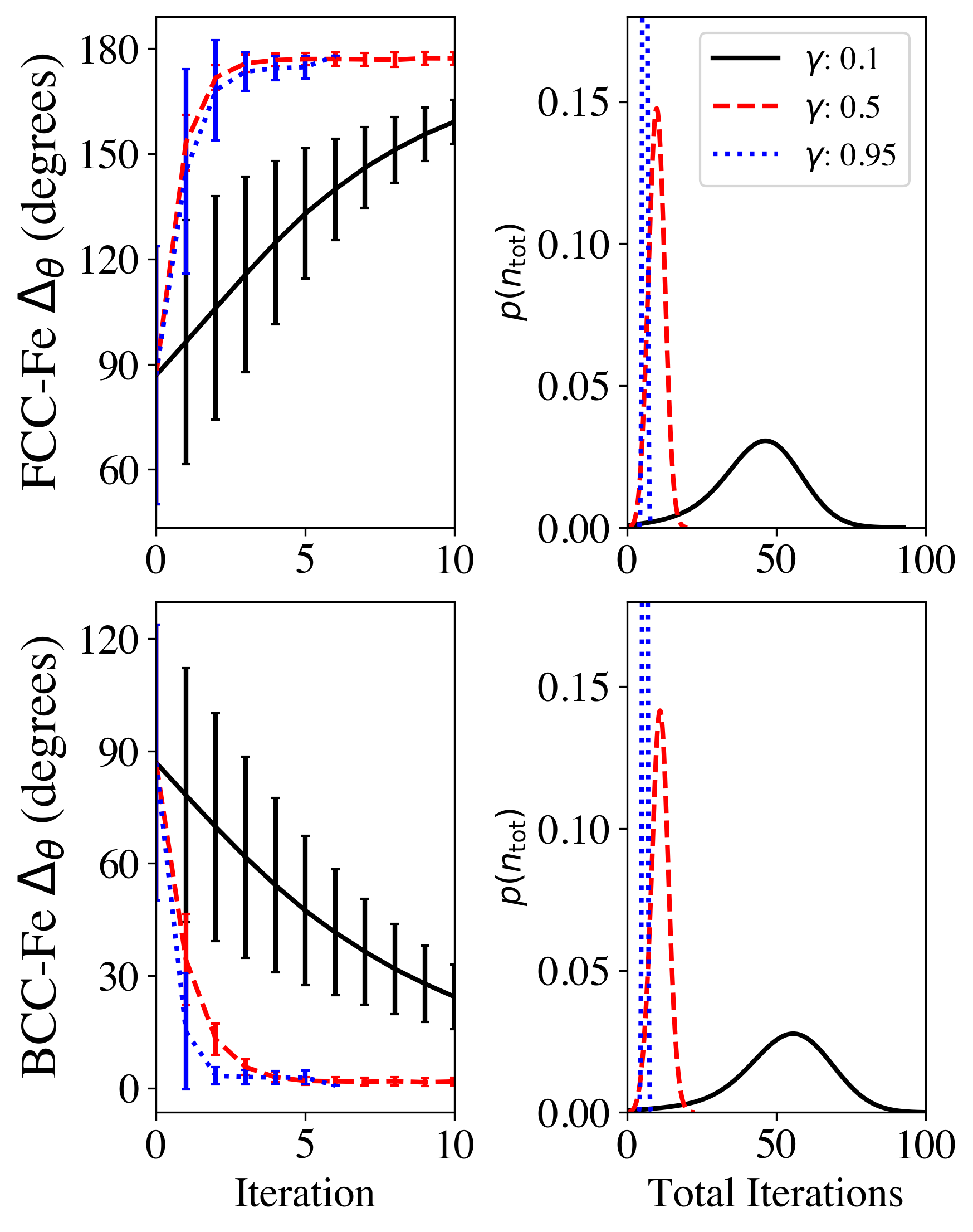}
\caption{Convergence of SELFE’s self-consistency loop. The evolution of the angle between nearest neighbor spins $\Delta_\theta$ as a function of convergence iterations is plotted for FCC-Fe (top) and BCC-Fe (bottom). The left panels display the mean and standard deviation of $\Delta_\theta$ for different values of the mixing parameter $\gamma$, based on 200 simulations with randomized initial spin configurations. The right panels show the probability of the total number of iterations $p(n_{\text{tot}})$ required to meet the Z-score convergence criterion for each $\gamma$ value. Lower $\gamma$ values converge more slowly but can show reduced variance, indicating greater stability during the convergence process.}
\label{fig:convergence}
\end{figure}

We now investigate the convergence of SELFE's self-consistency loop using the exchange parameters for FCC-Fe and BCC-Fe. We test convergence by running two hundred simulations starting from random spin configurations, $n=5$, and $T\ll T_{C/N}$. To validate the convergence properties of SELFE, we compute the angle between nearest neighbor spins $\Delta_\theta$ as a function of convergence iteration. We plot the mean and standard deviation of $\Delta_\theta$ in Fig. \ref{fig:convergence} for FCC-Fe (upper left) and BCC-Fe (lower left) for different values of mixing parameter $\gamma$. In addition, the total number of iterations required to satisfy the Z-score convergence criteria is provided on the right side of Fig. \ref{fig:convergence}.

In Fig. \ref{fig:convergence}, a $\Delta_\theta$ of $0^\circ$ and $180^\circ$ represent the FM and AFM states respectively. We see that every simulation converged to the expected magnetic ordering with lower mixing parameters converging more slowly. Interestingly, the $\gamma=0.95$ simulations appear to have a larger variance than the $\gamma=0.5$ simulations, likely due to the onset of instabilities as $\gamma \rightarrow 1$. The $\gamma=0.95$ simulations converge within the minimum possible number of iterations with $n=5$.

\begin{figure}[!htbp]
\centering
\includegraphics[scale=0.75]{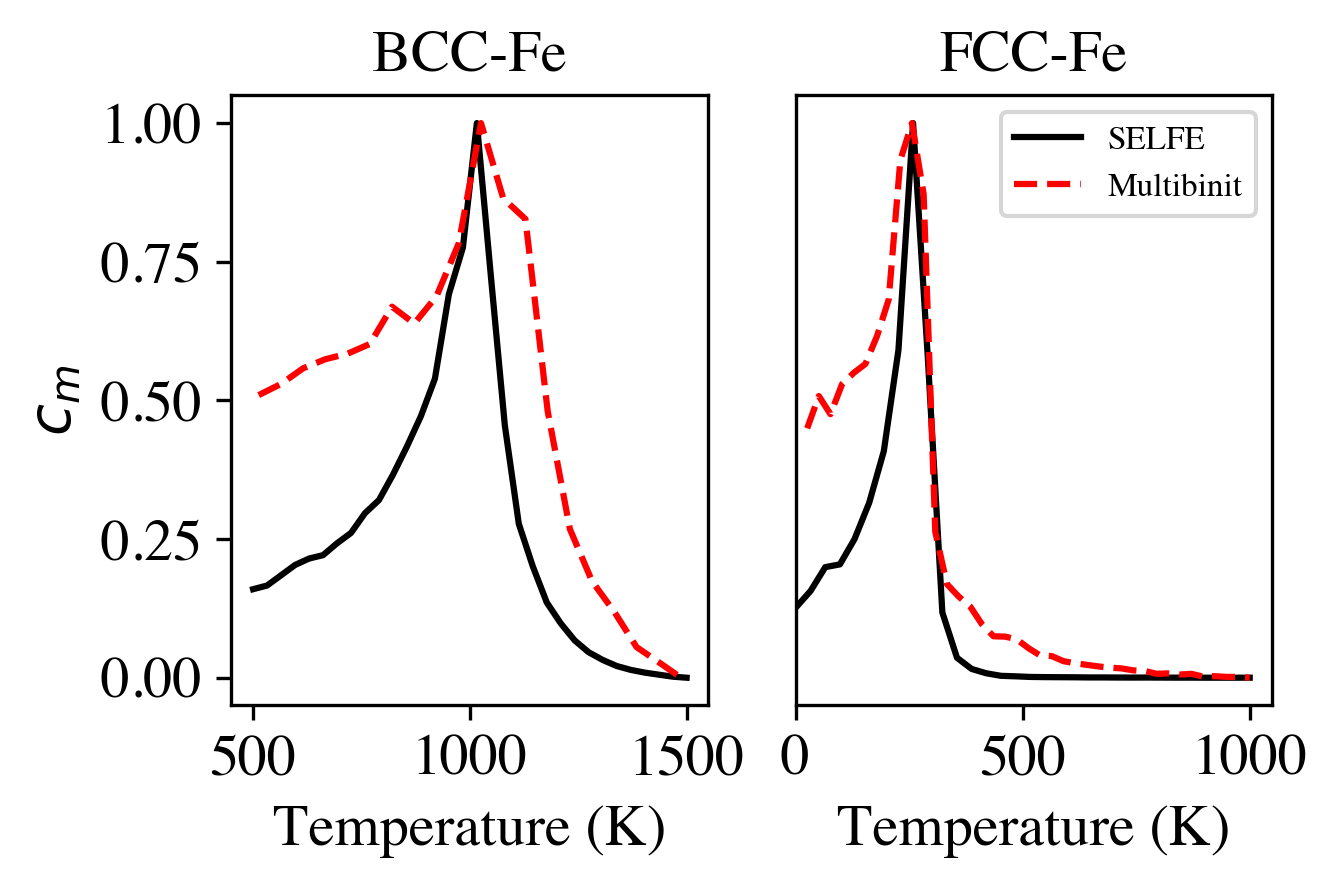}
\caption{Estimation of critical temperatures from normalized heat capacity curves. Comparison of SELFE (black solid lines) and Multibinit (red dashed lines) estimating $T_C$ of BCC-Fe and $T_N$ of FCC-Fe. The heat capacity $c_m$ curves are normalized between $[0, 1]$ to facilitate comparison. SELFE achieves excellent agreement with Multibinit, while requiring significantly fewer sampling steps, demonstrating its efficiency and accuracy for predicting critical temperatures.}
\label{fig:cv}
\end{figure}

Next, we explored the functional performance of our algorithm by calculating the heat capacity and magnetic phase transition temperatures for FM BCC-Fe and AFM FCC-Fe. We compare our algorithm to Multibinit results in Fig. \ref{fig:cv}.

As shown in Fig. \ref{fig:cv}, the normalized magnetic heat capacity curves produced by SELFE closely match those obtained using the Multibinit framework, accurately identifying the critical temperatures of both systems: \( T_C = 1{,}016 \, \mathrm{K} \) for BCC-Fe and \( T_N = 259 \, \mathrm{K} \) for FCC-Fe. In the case of FCC-Fe, further investigation is needed to explain why SELFE and Multibinit results agree but both differ from Ref.~\cite{haglof2021calphad} ($T_N=192$), which uses a different methodology and cell volume. The discrepancy may arise from differences in magnetic structure assumptions, lattice constraints, and the treatment of finite-temperature magnetic excitations, as well as the use of tetragonal distortions and approximations to the paramagnetic state in the CALPHAD framework. Despite differences from CALPHAD for antiferromagnetic FCC-Fe, SELFE consistently matched the accuracy of Multibinit for both materials while requiring only 350 self-consistent averaging steps. This is a distinct advantage when compared to Multibinit's 5{,}000 (BCC-Fe) and 2{,}000 (FCC-Fe) thermalization steps, followed by 50{,}000 (BCC-Fe) and 10{,}000 (FCC-Fe) averaging steps. This demonstrates a sample efficiency improvement of up to 142$\times$, highlighting SELFE's strength in producing accurate results with significantly reduced computational cost. We summarize this comparison in Table~\ref{tab:selfe_comparison}, which also includes results from Ref.~\cite{Mendive-Tapia2022} and experimental data. The table highlights SELFE's efficiency and accuracy in reproducing critical temperatures with orders-of-magnitude fewer simulation steps.

\begin{table}[ht]
\caption{\label{tab:selfe_comparison}%
Comparison of simulation parameters and estimated critical temperatures using SELFE, Multibinit, Ref.~\cite{Mendive-Tapia2022}, and experimental data. SELFE achieves comparable accuracy with significantly fewer simulation steps. Experimental values are taken from Refs.~\cite{he2022third, haglof2021calphad}. DFT requirements are reported as the product of the number of calculations and the number of atoms in the simulation cell. The superscript~$\textit{g}$ indicates a single Green's function calculation is required. The superscript~$\textit{c}$ indicates that calculations were performed under spin constraints.
}
\begin{ruledtabular}
\begin{tabular}{lcccc}
\textrm{Quantity} & \textrm{SELFE} & \textrm{Multibinit} & \textrm{Refs.~\cite{Mendive-Tapia2022, haglof2021calphad}} & \textrm{Exp} \\
\hline
\multicolumn{5}{l}{\textbf{BCC-Fe}} \\
Thermalization steps        & 1       & 5{,}000   &      &  \\
Averaging steps             & 350     & 50{,}000  &  10  &  \\
DFT Requirements  & 1x1$^\textit{g}$       & 1x1$^\textit{g}$         & 10x432$^\textit{c}$      &  \\
Estimated $T_C$ (K)         & 1{,}016 & 1{,}026   & 1{,}590  & 1{,}043 \\
\multicolumn{5}{l}{\textbf{FCC-Fe}} \\
Thermalization steps        & 1       & 2{,}000   &      &  \\
Averaging steps             & 350     & 10{,}000  &      &  \\
DFT Requirements  & 1x2$^\textit{g}$       & 1x2$^\textit{g}$    &      & \\
Estimated $T_N$ (K)         & 259     & 256       &  192  & 67 \\
\end{tabular}
\end{ruledtabular}
\end{table}

\section{Discussion} \label{Discussion}
The results above demonstrate that SELFE can accurately predict magnetothermal properties for systems with both ferromagnetic (FM) and antiferromagnetic (AFM) ordering using minimal computational resources and a compact implementation. Its predictions match the accuracy of Multibinit for both BCC-Fe and FCC-Fe and improves upon those reported in Ref.~\cite{Mendive-Tapia2022}.

For BCC-Fe, SELFE predicts \( T_C = 1{,}016 \, \mathrm{K} \), in excellent agreement with the experimental value of \( 1{,}043 \, \mathrm{K} \)~\cite{haglof2021calphad}. While SELFE, like many theoretical methods, overestimates the Néel temperature of FCC-Fe, its prediction of \( T_N = 259 \, \mathrm{K} \) closely matches the result from Multibinit and reflects the well-known trend of elevated theoretical estimates for AFM ordering in this system~\cite{haglof2021calphad, he2022third, li2021effects}.

In addition to accuracy, SELFE offers practical advantages. The method eliminates the need for free parameters, simplifying integration into automated high-throughput workflows. It also improves upon the framework proposed in Ref.~\cite{Mendive-Tapia2022} by incorporating self-consistency and requiring only a single DFT calculation and minimal sampling. When compared to established packages such as Multibinit, SELFE demonstrates efficiency gains, achieving convergence with orders-of-magnitude fewer thermalization and averaging steps while maintaining predictive accuracy.

The Z-score convergence criteria investigated in this study appears robust, effectively eliminating the need for human intervention. Throughout the investigation, no difficulties were encountered in obtaining self-consistency. Furthermore, the self-consistency loop reliably converged to the expected magnetic ordering for both FM and AFM systems. While the Z-score convergence criteria is intuitive from a statistical sense, simpler convergence criteria such as a threshold on $\Delta_i$ could be more straightforward to implement and interpret. Our informal investigations suggest that this is less robust compared to Z-score convergence criteria. With regards to mixing, we would recommend $\gamma=0.90$ as a reasonable default value because it balances convergence speed with stability. Going above this value occasionally caused convergence issues in the Z-score criteria, so for more complicated systems, we recommend lowering mixing to $\gamma = 0.5$.

Despite these promising results, more work is necessary to demonstrate SELFE's range of applicability to more general magnetic systems. Future work should focus on applying SELFE to a range of complex magnetic systems such as multiferroics \cite{mostovoy2024multiferroics} and spin glasses to evaluate its generalizability. Moreover, SELFE can be extended to predict a broader range of magnetothermal properties such as magnetocrystalline anisotropy \cite{daalderop1991magnetocrystalline, inzani2022manipulation}, magnetic susceptibility, and the Gibb's free energy expression. Future investigations may need to extend the second order Hamiltonian we consider to include higher order terms terms beyond the isotropic exchange interaction.

In this Letter, we introduced Sampled Effective Local Field Estimation (SELFE), a novel computational framework for predicting magnetothermal properties of materials from first principles. SELFE offers an efficient and accurate approach to calculating magnetothermal properties such as the Curie and N\'eel temperatures for systems exhibiting ferromagnetic (FM) and antiferromagnetic (AFM) ordering. Our results demonstrate strong agreement with established methods and experimental results while significantly reducing computational resource requirements and sampling steps. Furthermore, SELFE improves upon existing frameworks by incorporating two-body interactions, providing a fully automated, user-friendly implementation that is straightforward to integrate into high-throughput workflows.

\section*{Acknowledgments} \label{Acknowledgments}
This work is supported by the MITRE Independent Research and Development Program. We would like to thank Addis Fuhr from Oak Ridge National Laboratory and Luke Mauritsen from Quantum Strategix for fruitful conversations.

\bibliography{main.bib}

\clearpage
\onecolumngrid

\section*{Supplemental Material}

\input{si.tex}

\end{document}

%% file: si.tex
\fancyfoot[C]{\footnotesize Approved for Public Release; Distribution Unlimited. Public Release Case Number 25-1105. \\ {\footnotesize \copyright}\ 2025 The MITRE Corporation. All rights reserved.}
\fancyhead{}

\DeclareSIUnit\angstrom{\text {Å}}


\title{Supplemental Information for: Magnetothermal Properties with Sampled Effective Local Field Estimation}

\maketitle

\thispagestyle{fancy}
\section{Derivations}
\subsection*{I. Site-level Thermodynamic Distributions}
We now derive the site-level distribution over polar angles presented in the main text. The internal energy $U_{int}$ quantifies the magnetic contributions to the thermodynamic behavior of the material under consideration and can be written as:

\begin{equation} 
U_{int} = \int \prod_{  i \in  \{ m_1, \cdots, m_N \}  } [ d^3 \hat{\mathbf{e}_i} ]    P (  \{ \hat{\mathbf{e}}_{i} \} ) E_{int} (  \{ \hat{\mathbf{e}}_{i} \} )     
\end{equation}  

where $\hat{\mathbf{e}}_i$ is the magnetic moment for site $i$ and the product runs over all permutations of magnetic sites  $\{ m_1, \cdots, m_N \}$ in the lattice. We see that the integration is performed for each magnetic lattice site $i$ over a sphere ( $\int d^3  \hat{\mathbf{e}}_{i}  ( \cdots)  $ ).  $P (  \{ \hat{\mathbf{e}}_{i} \} )$  is the probability of being in configuration $\{ \hat{\mathbf{e}}_{i} \} $, requiring a method to evaluate it. We can express this as:

\begin{equation} \label{eqn:probability_appendix}
P  (  \{ \hat{\mathbf{e}}_{i} \} ) =  \frac{\exp (-\beta  E_{int} (  \{ \hat{\mathbf{e}}_{i} \} )    )}{Z}
\end{equation}

where with $\beta = \frac{1}{k_bT}$ and $Z$ as the partition function for the system. $E_{int} (  \{ \hat{\mathbf{e}}_{i} \} )$ is the internal energy for the magnetic configuration considered.     

Now we expand out the partition function $Z$: 
\begin{equation} \label{eqn: appendix_partitionfunc}
Z  =   \sum_{ m_{N} = 1}^{N_{\theta}^{N} \times N_\phi^{N}}    \cdots   \sum_{ m_{1} = 1}^{N_{\theta}^{1} \times N_\phi^{1}}  \exp (-\beta  E_{int} (  \{ \hat{\mathbf{e}}_{m_1, \cdots, m_N} \} )    )  
\end{equation}

where the summations run over microstates for each magnetic site $m_i$, parameterized by set of solid angles defining the orientation of the magnetic moment defined as $(\theta_i, \phi_i)$. Each summation index in Eq. (\ref{eqn: appendix_partitionfunc}) corresponds to a magnetic site and the summations run over the total number of possible solid angles in the discrete case, here being $N_\theta^i \times N_\phi^i$ where $N_\phi^i$ is the total number of discrete azimuthal angles and $N_\theta^i$ is the total number of polar angles that magnetic site $i$ can sit in.  Again, we attempt to make $Z$ continuous by replacing the summations with spherical integrals per magnetic site as shown:  
\begin{equation} 
Z =   \int \prod_{i=1}^{N_{sites}} [d^3 \hat{\mathbf{e}_i}]  \exp (-\beta  E_{int} (  \{ \hat{\mathbf{e}}_{m_1, \cdots, m_N} \} )    )   
\end{equation}

This motivates us to design a distribution $P_{tr}$  over magnetic configurations that is both analytically and numerically tractable.  One approach is to express the distribution in terms of an effective Hamiltonian to eliminate the need to calculate the internal energy $E_{int}$ explicitly. We can write out the following Hamiltonian:    

\begin{equation} 
\mathcal{H}_{tr} = -\sum_{i=1}^N \mathbf{h}_i^{(1)}  \cdot \hat{\mathbf{e}}_i  +  \sum_{i=1}^N \sum_{j=1}^N  h_{i,j}^{(2)}   \hat{\mathbf{e}}_i  \cdot \hat{\mathbf{e}}_j + \cdots 
\end{equation}  

For now, we will make the simplifying assumption to cut off the Hamiltonian to first order terms.  Hence, we only consider the expression $\mathcal{H}_{tr} = -\sum_{i=1}^N \mathbf{h}_i^{(1)}  \cdot \hat{\mathbf{e}}_i$ for now, corresponding to the Curie-Weiss expression in Eq. (1). It's corresponding probability distribution function is:  

\begin{equation} 
\begin{split} 
P_{tr} (  \{ \hat{\mathbf{e}}_{i} \} )  =  \frac{\exp (-\beta \mathcal{H}_{tr} ) }{Z_{tr}} \\
= \frac{\exp (\beta  \sum_{i=1}^N \mathbf{h}_i^{(1)}  \cdot \hat{\mathbf{e}}_i) }{Z_{tr}} \\
=  \frac{ \prod_{i=1}^N \exp (\beta   \mathbf{h}_i^{(1)}  \cdot \hat{\mathbf{e}}_i) }{Z_{tr}}  
\end{split}
\end{equation} 

Here $Z_{tr}$ is the trial partition function provided the trial first-order Hamiltonian. Note that in this first order case, the  terms $\mathbf{h}_i^{(1)}$ represent effective magnetic fields for each site in the lattice, described in further detail in the main text. In practice, we obtain these values from our self-consistent procedure coupled with the Green's function method described earlier. The trial partition function can be written as:
\begin{equation} 
Z_{tr} = \prod_{i=1}^N \int d^3 \hat{\mathbf{e}}_{i}  \exp( \beta \mathbf{h}_i^{(1)}  \cdot \hat{\mathbf{e}}_i  )      
\end{equation}  
where $N$ is the total number of magnetic sites. 

Evaluating the partition function, we will assume that $\mathbf{h}_n^{(1)}$ is oriented along the $[0,0,1]$ crystal orientation, hence $\mathbf{h}_n^{(1)} = [0,0,h_n^{(1)}]$. $Z_{tr}$ then can be written as:
\begin{equation} 
\begin{split}
Z_{tr} =  \prod_{i=1}^N \int_0^{2\pi} \int_0^{\pi}  d \phi_i d\theta_i \sin \theta_i   \exp( \beta h_i^{(1)}   \cos \theta_i       ) \\  
= \prod_{i=1}^N  4\pi  \frac{\sinh ( \beta h_i^{(1)}) }{\beta h_i^{(1)}}  
\end{split} 
\end{equation}  
 
This now allows us to write out the probability distribution over magnetic configurations, which was our original goal:   

\begin{equation} 
\begin{split} 
P_{tr} (  \{ \hat{\mathbf{e}}_{i} \} ) =  \frac{ \prod_{i=1}^N \exp (\beta   \mathbf{h}_i^{(1) }  \cdot \hat{\mathbf{e}}_i )  }{Z_{tr}}  \\  
= \frac{\prod_{i=1}^N    \exp (\beta   h_i^{(1) }  \cos \theta_i)    }{  \prod_{i=1}^N  4\pi  \frac{\sinh ( \beta h_i^{(1)}) }{\beta h_i^{(1)}}  } \\
= \prod_{i=1}^N  \frac{ \beta h_i^{(1)} \exp (\beta   h_i^{(1) }  \cos \theta_i)    }{ \sinh ( \beta h_i^{(1)})}  
\end{split} 
\end{equation} 

In practice, we can assign the order parameter $h_i^{(1)}$ to vary according to the desired magnetic phase and to be unique for each magnetic species in the lattice. $h_i^{(1)} \rightarrow 0$ corresponds to a disordered phase while $h_i^{(1)} \rightarrow \infty $ corresponds to a magnetically order phase. We will define a new probability $P_n ( \hat{\mathbf{e}}_{n} ; \theta_n )$ for this first order case such that $P_{tr} (  \{ \hat{\mathbf{e}}_{i} \} ) = \prod_{i=1}^N P_i ( \hat{\mathbf{e}}_{i} ; \theta_i ) $ (note that we take $ \hat{\mathbf{e}}_n = (\sin \theta_n \cos \phi_n , \sin \theta_n \sin \phi_n, \cos \theta_n) $).

\begin{equation} 
 P_n ( \hat{\mathbf{e}}_{n} ; \theta_n ) =   \frac{ \beta h_n^{(1)} \exp (\beta   h_n^{(1) }  \cos \theta_n)    }{ \sinh ( \beta h_n^{(1)})} 
 \end{equation} 
This provides us with a physically-grounded sampling procedure that we use to execute our SELFE approach.

\section{DFT Input Files}
\subsubsection{BCC Fe}
\begin{verbatim}
SystemName	BCCFe
SystemLabel	siesta
CDF.Compress	9
CDF.Save	True
MaxSCFIteration	60
SCF.DM.Tolerance	0.0001
SCF.EDM.Tolerance	1e-2 eV
SCF.H.Tolerance	1e-3 eV
SCF.Mixer.History	16
SCF.Mixer.Method	Pulay
SCF.Mixer.Spin	spinor
SCF.Mixer.Weight	0.4
SCF.Spin.Fix	True
SaveHS	True
Write.DMHS.Netcdf	True
SCFMustConverge	True
MD.TypeOfRun          CG
MD.NumCGSteps         100
MD.VariableCell       T
MD.MaxForceTol        0.1 eV/Ang
MD.MaxStressTol       0.1 GPa
MD.TargetPressure     0.0 GPa
Spin     	COLLINEAR
XC.functional	GGA
XC.authors	PBE
MeshCutoff	2721.1386024367243	eV
PAO.EnergyShift	0.1	eV
NumberOfSpecies	1
NumberOfAtoms	1
%block ChemicalSpecieslabel
    1 26 Fe.1
%endblock ChemicalSpecieslabel
%block PAO.BasisSizes
    Fe.1	DZP
%endblock PAO.BasisSizes
LatticeConstant	1.0 Ang
%block LatticeVectors
    2.466540770000000     0.000000000000000     0.000000000000000
    -0.822180256484402     2.325476939451225     0.000000000000000
    -0.822180256484402     -1.162738469338971     2.013922105702875
%endblock LatticeVectors
AtomicCoordinatesFormat  Ang
%block AtomicCoordinatesAndAtomicSpecies
     0.000000000      0.000000000      0.000000000 1
%endblock AtomicCoordinatesAndAtomicSpecies
%block DM.InitSpin
    1 3.00000000000000
%endblock DM.InitSpin
DM.UseSaveDM	True
#KPoint grid
%block kgrid_Monkhorst_Pack
     14       0       0  0.0
     0       14       0  0.0
     0       0       14  0.0
%endblock kgrid_Monkhorst_Pack
# Output options
Charge.Mulliken end
Charge.Mulliken.Format 1
SaveRho T
WriteDenchar T
\end{verbatim}

\subsubsection{FCC Fe}
\begin{verbatim}
SystemName	FCC Fe
SystemLabel	siesta
CDF.Compress	9
CDF.Save	True
MaxSCFIteration	150
SCF.DM.Tolerance	0.0001
SCF.EDM.Tolerance	1e-2 eV
SCF.H.Tolerance	1e-3 eV
SCF.Mixer.History	16
SCF.Mixer.Method	Pulay
SCF.Mixer.Spin	spinor
SCF.Mixer.Weight	0.4
SCF.Spin.Fix	True
SaveHS	True
Write.DMHS.Netcdf	True
SCFMustConverge	True
MD.TypeOfRun          CG
MD.NumCGSteps         100
MD.VariableCell       T
MD.MaxForceTol        0.1 eV/Ang
MD.MaxStressTol       0.1 GPa
MD.TargetPressure     0.0 GPa
# Spin options
Spin     	COLLINEAR
%block DM.InitSpin
  1 +
  2 -
  3 +
  4 -
%endblock DM.InitSpin
XC.functional	GGA
XC.authors	PBE
MeshCutoff	2721.1386024367243	eV
PAO.EnergyShift	0.1	eV
NumberOfSpecies	1
NumberOfAtoms	4
%block ChemicalSpecieslabel
    1 26 Fe.1
%endblock ChemicalSpecieslabel
%block PAO.BasisSizes
    Fe.1	DZP
%endblock PAO.BasisSizes
# Lattice vectors
LatticeConstant  1.0 Ang
%block LatticeVectors
   3.6555511703502481   0.0000000000000000   0.0000000000000002
   0.0000000000000006   3.6555511703502481   0.0000000000000002
   0.0000000000000000   0.0000000000000000   3.6555511703502481
%endblock LatticeVectors
#Atomic coordinates
AtomicCoordinatesFormat  Fractional
%block AtomicCoordinatesAndAtomicSpecies
  0.0000000000000000  0.0000000000000000  0.0000000000000000  1
  0.0000000000000000  0.5000000000000000  0.5000000000000000  1
  0.5000000000000000  0.0000000000000000  0.5000000000000000  1
  0.5000000000000000  0.5000000000000000  0.0000000000000000  1
%endblock AtomicCoordinatesAndAtomicSpecies
DM.UseSaveDM	True
#KPoint grid
%block kgrid_Monkhorst_Pack
     14       0       0  0.0
     0       14       0  0.0
     0       0       14  0.0
%endblock kgrid_Monkhorst_Pack
# Output options
Charge.Mulliken end
Charge.Mulliken.Format 1
SaveRho T
WriteDenchar T
\end{verbatim}

\section{Rotation Matrices}
The polar coordinates of $\mathbf{\hat{e}}_i$ are initially expressed in the local coordinate frame of each magnetic site so that $(\theta_i, \phi_i)$ are angles measured from $\mathbf{\bar{h}}_i$ as the positive z-direction in the local frame. To express the direction of $\mathbf{\hat{e}}_i$ in the global coordinate frame of the crystal, we apply a rotation matrix ($R_h^0$) that translates the local coordinates (denoted as frame h) onto the global coordinates (denoted as frame 0). $R_h^0$ is composed of two successive rotations that bring the global $z$ axis into alignment with the local $\mathbf{\bar{h}}_i$: first about the global \( y \)-axis by an angle \( \beta_i \), and then about the global \( z \)-axis by an angle \( \alpha_i \). The combined rotation matrix is given by:

\begin{equation}\label{eq:rotation}
\begin{split}
R_h^0 &= R_{z,\alpha_i}\, R_{y,\beta_i} \\
&= \begin{pmatrix}
\cos{\alpha_i} & -\sin{\alpha_i} & 0 \\
\sin{\alpha_i} & \cos{\alpha_i} & 0 \\
0 & 0 & 1
\end{pmatrix}
\begin{pmatrix}
\cos{\beta_i} & 0 & \sin{\beta_i} \\
0 & 1 & 0 \\
-\sin{\beta_i} & 0 & \cos{\beta_i} \\
\end{pmatrix} \\
&= \begin{pmatrix}
\cos \alpha_i \cos \beta_i & -\sin \alpha_i & \cos \alpha_i \sin \beta_i \\
\sin \alpha_i \cos \beta_i & \cos \alpha_i & \sin \alpha_i \sin \beta_i \\
-\sin \beta_i & 0 & \cos \beta_i
\end{pmatrix},
\end{split}
\end{equation}

where the angles \( \alpha_i \) and \( \beta_i \) are determined by the orientation of the local magnetic field vector \( \mathbf{\bar{h}}_i = (h_{ix}, h_{iy}, h_{iz}) \). These angles are given by:

\begin{align}
\alpha_i &= \arctan2(h_{iy}, h_{ix}), \\
\beta_i  &= \arccos\left(\frac{h_{iz}}{|\mathbf{\bar{h}}_i|}\right), \quad \text{with} \quad |\mathbf{\bar{h}}_i| = \sqrt{h_{ix}^2 + h_{iy}^2 + h_{iz}^2}.
\end{align}

The function \( \arctan2 \) ensures correct quadrant determination for \( \alpha_i \). The rotation matrix \( R_h^0 \) is then used to transform magnetic moment $\mathbf{\hat{e}}_i$ from the local frame aligned with \( \mathbf{\bar{h}}_i \) into the global reference frame.


%% file: main.bbl
\begin{thebibliography}{40}%
\makeatletter
\providecommand \@ifxundefined [1]{%
 \@ifx{#1\undefined}
}%
\providecommand \@ifnum [1]{%
 \ifnum #1\expandafter \@firstoftwo
 \else \expandafter \@secondoftwo
 \fi
}%
\providecommand \@ifx [1]{%
 \ifx #1\expandafter \@firstoftwo
 \else \expandafter \@secondoftwo
 \fi
}%
\providecommand \natexlab [1]{#1}%
\providecommand \enquote  [1]{``#1''}%
\providecommand \bibnamefont  [1]{#1}%
\providecommand \bibfnamefont [1]{#1}%
\providecommand \citenamefont [1]{#1}%
\providecommand \href@noop [0]{\@secondoftwo}%
\providecommand \href [0]{\begingroup \@sanitize@url \@href}%
\providecommand \@href[1]{\@@startlink{#1}\@@href}%
\providecommand \@@href[1]{\endgroup#1\@@endlink}%
\providecommand \@sanitize@url [0]{\catcode `\\12\catcode `\$12\catcode
  `\&12\catcode `\#12\catcode `\^12\catcode `\_12\catcode `\%12\relax}%
\providecommand \@@startlink[1]{}%
\providecommand \@@endlink[0]{}%
\providecommand \url  [0]{\begingroup\@sanitize@url \@url }%
\providecommand \@url [1]{\endgroup\@href {#1}{\urlprefix }}%
\providecommand \urlprefix  [0]{URL }%
\providecommand \Eprint [0]{\href }%
\providecommand \doibase [0]{https://doi.org/}%
\providecommand \selectlanguage [0]{\@gobble}%
\providecommand \bibinfo  [0]{\@secondoftwo}%
\providecommand \bibfield  [0]{\@secondoftwo}%
\providecommand \translation [1]{[#1]}%
\providecommand \BibitemOpen [0]{}%
\providecommand \bibitemStop [0]{}%
\providecommand \bibitemNoStop [0]{.\EOS\space}%
\providecommand \EOS [0]{\spacefactor3000\relax}%
\providecommand \BibitemShut  [1]{\csname bibitem#1\endcsname}%
\let\auto@bib@innerbib\@empty
\bibitem [{\citenamefont {Gschneidner}\ \emph {et~al.}(2003)\citenamefont
  {Gschneidner}, \citenamefont {Pecharsky},\ and\ \citenamefont
  {Pecharsky}}]{Gschneidner2003}%
  \BibitemOpen
  \bibfield  {author} {\bibinfo {author} {\bibfnamefont {K.~A.}\ \bibnamefont
  {Gschneidner}}, \bibinfo {author} {\bibfnamefont {A.~O.}\ \bibnamefont
  {Pecharsky}},\ and\ \bibinfo {author} {\bibfnamefont {V.~K.}\ \bibnamefont
  {Pecharsky}},\ }\bibinfo {title} {Low temperature cryocooler regenerator
  materials},\ in\ \href {https://doi.org/10.1007/0-306-47919-2_61} {\emph
  {\bibinfo {booktitle} {Cryocoolers 12}}},\ \bibinfo {editor} {edited by\
  \bibinfo {editor} {\bibfnamefont {R.~G.}\ \bibnamefont {Ross}}}\ (\bibinfo
  {publisher} {Springer US},\ \bibinfo {address} {Boston, MA},\ \bibinfo {year}
  {2003})\ pp.\ \bibinfo {pages} {457--465}\BibitemShut {NoStop}%
\bibitem [{\citenamefont {Numazawa}\ \emph {et~al.}(2002)\citenamefont
  {Numazawa}, \citenamefont {Arai}, \citenamefont {Sato}, \citenamefont
  {Fujimoto}, \citenamefont {Oodo}, \citenamefont {Rang},\ and\ \citenamefont
  {Yanagitani}}]{Numazawa2002}%
  \BibitemOpen
  \bibfield  {author} {\bibinfo {author} {\bibfnamefont {T.}~\bibnamefont
  {Numazawa}}, \bibinfo {author} {\bibfnamefont {O.}~\bibnamefont {Arai}},
  \bibinfo {author} {\bibfnamefont {A.}~\bibnamefont {Sato}}, \bibinfo {author}
  {\bibfnamefont {S.}~\bibnamefont {Fujimoto}}, \bibinfo {author}
  {\bibfnamefont {T.}~\bibnamefont {Oodo}}, \bibinfo {author} {\bibfnamefont
  {Y.~M.}\ \bibnamefont {Rang}},\ and\ \bibinfo {author} {\bibfnamefont
  {T.}~\bibnamefont {Yanagitani}},\ }\bibinfo {title} {New regenerator material
  for sub-4 k cryocoolers},\ in\ \href
  {https://doi.org/10.1007/0-306-47112-4_59} {\emph {\bibinfo {booktitle}
  {Cryocoolers 11}}},\ \bibinfo {editor} {edited by\ \bibinfo {editor}
  {\bibfnamefont {R.~G.}\ \bibnamefont {Ross}}}\ (\bibinfo  {publisher}
  {Springer US},\ \bibinfo {address} {Boston, MA},\ \bibinfo {year} {2002})\
  pp.\ \bibinfo {pages} {465--473}\BibitemShut {NoStop}%
\bibitem [{\citenamefont {Manipatruni}\ \emph {et~al.}(2019)\citenamefont
  {Manipatruni}, \citenamefont {Nikonov}, \citenamefont {Lin}, \citenamefont
  {Gosavi}, \citenamefont {Liu}, \citenamefont {Prasad}, \citenamefont {Huang},
  \citenamefont {Bonturim}, \citenamefont {Ramesh},\ and\ \citenamefont
  {Young}}]{manipatruni2019scalable}%
  \BibitemOpen
  \bibfield  {author} {\bibinfo {author} {\bibfnamefont {S.}~\bibnamefont
  {Manipatruni}}, \bibinfo {author} {\bibfnamefont {D.~E.}\ \bibnamefont
  {Nikonov}}, \bibinfo {author} {\bibfnamefont {C.-C.}\ \bibnamefont {Lin}},
  \bibinfo {author} {\bibfnamefont {T.~A.}\ \bibnamefont {Gosavi}}, \bibinfo
  {author} {\bibfnamefont {H.}~\bibnamefont {Liu}}, \bibinfo {author}
  {\bibfnamefont {B.}~\bibnamefont {Prasad}}, \bibinfo {author} {\bibfnamefont
  {Y.-L.}\ \bibnamefont {Huang}}, \bibinfo {author} {\bibfnamefont
  {E.}~\bibnamefont {Bonturim}}, \bibinfo {author} {\bibfnamefont
  {R.}~\bibnamefont {Ramesh}},\ and\ \bibinfo {author} {\bibfnamefont {I.~A.}\
  \bibnamefont {Young}},\ }\bibfield  {title} {\bibinfo {title} {Scalable
  energy-efficient magnetoelectric spin-orbit logic},\ }\href@noop {}
  {\bibfield  {journal} {\bibinfo  {journal} {Nature}\ }\textbf {\bibinfo
  {volume} {565}},\ \bibinfo {pages} {35} (\bibinfo {year} {2019})}\BibitemShut
  {NoStop}%
\bibitem [{\citenamefont {Han}\ and\ \citenamefont
  {Zhang}(2022)}]{han2022cryogenic}%
  \BibitemOpen
  \bibfield  {author} {\bibinfo {author} {\bibfnamefont {Y.}~\bibnamefont
  {Han}}\ and\ \bibinfo {author} {\bibfnamefont {A.}~\bibnamefont {Zhang}},\
  }\bibfield  {title} {\bibinfo {title} {Cryogenic technology for infrared
  detection in space},\ }\href@noop {} {\bibfield  {journal} {\bibinfo
  {journal} {Scientific reports}\ }\textbf {\bibinfo {volume} {12}},\ \bibinfo
  {pages} {2349} (\bibinfo {year} {2022})}\BibitemShut {NoStop}%
\bibitem [{\citenamefont {Johnson}\ \emph {et~al.}(2022)\citenamefont
  {Johnson}, \citenamefont {Pilat}, \citenamefont {Qualls}, \citenamefont
  {Colby}, \citenamefont {Fisher}, \citenamefont {Melcer},\ and\ \citenamefont
  {Runkles}}]{Johnson2022}%
  \BibitemOpen
  \bibfield  {author} {\bibinfo {author} {\bibfnamefont {E.}~\bibnamefont
  {Johnson}}, \bibinfo {author} {\bibfnamefont {F.}~\bibnamefont {Pilat}},
  \bibinfo {author} {\bibfnamefont {H.}~\bibnamefont {Qualls}}, \bibinfo
  {author} {\bibfnamefont {E.}~\bibnamefont {Colby}}, \bibinfo {author}
  {\bibfnamefont {K.}~\bibnamefont {Fisher}}, \bibinfo {author} {\bibfnamefont
  {N.}~\bibnamefont {Melcer}},\ and\ \bibinfo {author} {\bibfnamefont
  {K.}~\bibnamefont {Runkles}},\ }\href {https://doi.org/10.2172/1863553}
  {\emph {\bibinfo {title} {{Supply Chain Risk Mitigation for Scientific
  Facilities and Tools}}}},\ \bibinfo {type} {Tech. Rep.}\ \bibinfo {number}
  {November}\ (\bibinfo  {institution} {USDOE Office of Science (SC) (United
  States)},\ \bibinfo {year} {2022})\BibitemShut {NoStop}%
\bibitem [{\citenamefont {Rajan}(1983)}]{Rajan1983}%
  \BibitemOpen
  \bibfield  {author} {\bibinfo {author} {\bibfnamefont {V.~T.}\ \bibnamefont
  {Rajan}},\ }\bibfield  {title} {\bibinfo {title} {{Magnetic Susceptibility
  and Specific Heat of the Coqblin-Schrieffer Model}},\ }\href
  {https://doi.org/10.1103/PhysRevLett.51.308} {\bibfield  {journal} {\bibinfo
  {journal} {Physical Review Letters}\ }\textbf {\bibinfo {volume} {51}},\
  \bibinfo {pages} {308} (\bibinfo {year} {1983})}\BibitemShut {NoStop}%
\bibitem [{\citenamefont {Ghivelder}\ \emph {et~al.}(1999)\citenamefont
  {Ghivelder}, \citenamefont {{Abrego Castillo}}, \citenamefont {Gusm{\~{a}}o},
  \citenamefont {Alonso},\ and\ \citenamefont {Cohen}}]{Ghivelder1999}%
  \BibitemOpen
  \bibfield  {author} {\bibinfo {author} {\bibfnamefont {L.}~\bibnamefont
  {Ghivelder}}, \bibinfo {author} {\bibfnamefont {I.}~\bibnamefont {{Abrego
  Castillo}}}, \bibinfo {author} {\bibfnamefont {M.~A.}\ \bibnamefont
  {Gusm{\~{a}}o}}, \bibinfo {author} {\bibfnamefont {J.~A.}\ \bibnamefont
  {Alonso}},\ and\ \bibinfo {author} {\bibfnamefont {L.~F.}\ \bibnamefont
  {Cohen}},\ }\bibfield  {title} {\bibinfo {title} {{Specific heat and magnetic
  order in LaMnO}},\ }\href {https://doi.org/10.1103/PhysRevB.60.12184}
  {\bibfield  {journal} {\bibinfo  {journal} {Physical Review B}\ }\textbf
  {\bibinfo {volume} {60}},\ \bibinfo {pages} {12184} (\bibinfo {year}
  {1999})},\ \Eprint {https://arxiv.org/abs/9904232v1} {arXiv:9904232v1
  [arXiv:cond-mat]} \BibitemShut {NoStop}%
\bibitem [{\citenamefont {Ohno}\ \emph {et~al.}(1997)\citenamefont {Ohno},
  \citenamefont {Mauri},\ and\ \citenamefont {Louie}}]{Ohno1997}%
  \BibitemOpen
  \bibfield  {author} {\bibinfo {author} {\bibfnamefont {K.}~\bibnamefont
  {Ohno}}, \bibinfo {author} {\bibfnamefont {F.}~\bibnamefont {Mauri}},\ and\
  \bibinfo {author} {\bibfnamefont {S.~G.}\ \bibnamefont {Louie}},\ }\bibfield
  {title} {\bibinfo {title} {{Magnetic susceptibility of semiconductors by an
  all-electron first-principles approach}},\ }\href
  {https://doi.org/10.1103/PhysRevB.56.1009} {\bibfield  {journal} {\bibinfo
  {journal} {Physical Review B}\ }\textbf {\bibinfo {volume} {56}},\ \bibinfo
  {pages} {1009} (\bibinfo {year} {1997})}\BibitemShut {NoStop}%
\bibitem [{\citenamefont {Mauri}\ and\ \citenamefont
  {Louie}(1996)}]{Mauri1996}%
  \BibitemOpen
  \bibfield  {author} {\bibinfo {author} {\bibfnamefont {F.}~\bibnamefont
  {Mauri}}\ and\ \bibinfo {author} {\bibfnamefont {S.~G.}\ \bibnamefont
  {Louie}},\ }\bibfield  {title} {\bibinfo {title} {{Magnetic Susceptibility of
  Insulators from First Principles}},\ }\href
  {https://doi.org/10.1103/PhysRevLett.76.4246} {\bibfield  {journal} {\bibinfo
   {journal} {Physical Review Letters}\ }\textbf {\bibinfo {volume} {76}},\
  \bibinfo {pages} {4246} (\bibinfo {year} {1996})},\ \Eprint
  {https://arxiv.org/abs/9604008} {arXiv:9604008 [mtrl-th]} \BibitemShut
  {NoStop}%
\bibitem [{\citenamefont {Daalderop}\ \emph {et~al.}(1990)\citenamefont
  {Daalderop}, \citenamefont {Kelly},\ and\ \citenamefont
  {Schuurmans}}]{Daalderop1990}%
  \BibitemOpen
  \bibfield  {author} {\bibinfo {author} {\bibfnamefont {G.~H.~O.}\
  \bibnamefont {Daalderop}}, \bibinfo {author} {\bibfnamefont {P.~J.}\
  \bibnamefont {Kelly}},\ and\ \bibinfo {author} {\bibfnamefont {M.~F.~H.}\
  \bibnamefont {Schuurmans}},\ }\bibfield  {title} {\bibinfo {title}
  {{First-principles calculation of the magnetocrystalline anisotropy energy of
  iron, cobalt, and nickel}},\ }\href
  {https://doi.org/10.1103/PhysRevB.41.11919} {\bibfield  {journal} {\bibinfo
  {journal} {Physical Review B}\ }\textbf {\bibinfo {volume} {41}},\ \bibinfo
  {pages} {11919} (\bibinfo {year} {1990})}\BibitemShut {NoStop}%
\bibitem [{\citenamefont {Cooley}\ \emph {et~al.}(2023)\citenamefont {Cooley},
  \citenamefont {Dairaghi}, \citenamefont {Moore}, \citenamefont {Horton},
  \citenamefont {Schueller}, \citenamefont {Persson},\ and\ \citenamefont
  {Seshadri}}]{Cooley2023}%
  \BibitemOpen
  \bibfield  {author} {\bibinfo {author} {\bibfnamefont {J.~A.}\ \bibnamefont
  {Cooley}}, \bibinfo {author} {\bibfnamefont {G.}~\bibnamefont {Dairaghi}},
  \bibinfo {author} {\bibfnamefont {G.~C.}\ \bibnamefont {Moore}}, \bibinfo
  {author} {\bibfnamefont {M.~K.}\ \bibnamefont {Horton}}, \bibinfo {author}
  {\bibfnamefont {E.~C.}\ \bibnamefont {Schueller}}, \bibinfo {author}
  {\bibfnamefont {K.~A.}\ \bibnamefont {Persson}},\ and\ \bibinfo {author}
  {\bibfnamefont {R.}~\bibnamefont {Seshadri}},\ }\bibfield  {title} {\bibinfo
  {title} {Magnetism and magnetocaloric properties of co $ \_ $\{$1-x$\}$ $ mn
  $ \_x $ cr $ \_2 $ o $ \_4$},\ }\href@noop {} {\bibfield  {journal} {\bibinfo
   {journal} {arXiv preprint arXiv:2309.16168}\ } (\bibinfo {year}
  {2023})}\BibitemShut {NoStop}%
\bibitem [{\citenamefont {Shang}\ \emph {et~al.}(2010)\citenamefont {Shang},
  \citenamefont {Saal}, \citenamefont {Mei}, \citenamefont {Wang},\ and\
  \citenamefont {Liu}}]{Shang2010}%
  \BibitemOpen
  \bibfield  {author} {\bibinfo {author} {\bibfnamefont {S.-L.}\ \bibnamefont
  {Shang}}, \bibinfo {author} {\bibfnamefont {J.~E.}\ \bibnamefont {Saal}},
  \bibinfo {author} {\bibfnamefont {Z.-G.}\ \bibnamefont {Mei}}, \bibinfo
  {author} {\bibfnamefont {Y.}~\bibnamefont {Wang}},\ and\ \bibinfo {author}
  {\bibfnamefont {Z.-K.}\ \bibnamefont {Liu}},\ }\bibfield  {title} {\bibinfo
  {title} {{Magnetic thermodynamics of fcc Ni from first-principles partition
  function approach}},\ }\bibfield  {journal} {\bibinfo  {journal} {Journal of
  Applied Physics}\ }\textbf {\bibinfo {volume} {108}},\ \href
  {https://doi.org/10.1063/1.3524480} {10.1063/1.3524480} (\bibinfo {year}
  {2010})\BibitemShut {NoStop}%
\bibitem [{\citenamefont {Kvashnin}\ \emph {et~al.}(2015)\citenamefont
  {Kvashnin}, \citenamefont {Gr{\aa}n{\"{a}}s}, \citenamefont {{Di Marco}},
  \citenamefont {Katsnelson}, \citenamefont {Lichtenstein},\ and\ \citenamefont
  {Eriksson}}]{Kvashnin2015}%
  \BibitemOpen
  \bibfield  {author} {\bibinfo {author} {\bibfnamefont {Y.~O.}\ \bibnamefont
  {Kvashnin}}, \bibinfo {author} {\bibfnamefont {O.}~\bibnamefont
  {Gr{\aa}n{\"{a}}s}}, \bibinfo {author} {\bibfnamefont {I.}~\bibnamefont {{Di
  Marco}}}, \bibinfo {author} {\bibfnamefont {M.~I.}\ \bibnamefont
  {Katsnelson}}, \bibinfo {author} {\bibfnamefont {A.~I.}\ \bibnamefont
  {Lichtenstein}},\ and\ \bibinfo {author} {\bibfnamefont {O.}~\bibnamefont
  {Eriksson}},\ }\bibfield  {title} {\bibinfo {title} {{Exchange parameters of
  strongly correlated materials: Extraction from spin-polarized density
  functional theory plus dynamical mean-field theory}},\ }\href
  {https://doi.org/10.1103/PhysRevB.91.125133} {\bibfield  {journal} {\bibinfo
  {journal} {Physical Review B}\ }\textbf {\bibinfo {volume} {91}},\ \bibinfo
  {pages} {125133} (\bibinfo {year} {2015})},\ \Eprint
  {https://arxiv.org/abs/1503.02864} {arXiv:1503.02864} \BibitemShut {NoStop}%
\bibitem [{\citenamefont {Jain}\ \emph {et~al.}(2020)\citenamefont {Jain},
  \citenamefont {Montoya}, \citenamefont {Dwaraknath}, \citenamefont
  {Zimmermann}, \citenamefont {Dagdelen}, \citenamefont {Horton}, \citenamefont
  {Huck}, \citenamefont {Winston}, \citenamefont {Cholia}, \citenamefont {Ong}
  \emph {et~al.}}]{jain2020materials}%
  \BibitemOpen
  \bibfield  {author} {\bibinfo {author} {\bibfnamefont {A.}~\bibnamefont
  {Jain}}, \bibinfo {author} {\bibfnamefont {J.}~\bibnamefont {Montoya}},
  \bibinfo {author} {\bibfnamefont {S.}~\bibnamefont {Dwaraknath}}, \bibinfo
  {author} {\bibfnamefont {N.~E.}\ \bibnamefont {Zimmermann}}, \bibinfo
  {author} {\bibfnamefont {J.}~\bibnamefont {Dagdelen}}, \bibinfo {author}
  {\bibfnamefont {M.}~\bibnamefont {Horton}}, \bibinfo {author} {\bibfnamefont
  {P.}~\bibnamefont {Huck}}, \bibinfo {author} {\bibfnamefont {D.}~\bibnamefont
  {Winston}}, \bibinfo {author} {\bibfnamefont {S.}~\bibnamefont {Cholia}},
  \bibinfo {author} {\bibfnamefont {S.~P.}\ \bibnamefont {Ong}}, \emph
  {et~al.},\ }\bibfield  {title} {\bibinfo {title} {The materials project:
  Accelerating materials design through theory-driven data and tools},\
  }\href@noop {} {\bibfield  {journal} {\bibinfo  {journal} {Handbook of
  Materials Modeling: Methods: Theory and Modeling}\ ,\ \bibinfo {pages}
  {1751}} (\bibinfo {year} {2020})}\BibitemShut {NoStop}%
\bibitem [{\citenamefont {Togo}(2023)}]{togo2023first}%
  \BibitemOpen
  \bibfield  {author} {\bibinfo {author} {\bibfnamefont {A.}~\bibnamefont
  {Togo}},\ }\bibfield  {title} {\bibinfo {title} {First-principles phonon
  calculations with phonopy and phono3py},\ }\href@noop {} {\bibfield
  {journal} {\bibinfo  {journal} {Journal of the Physical Society of Japan}\
  }\textbf {\bibinfo {volume} {92}},\ \bibinfo {pages} {012001} (\bibinfo
  {year} {2023})}\BibitemShut {NoStop}%
\bibitem [{\citenamefont {S{\o}nden{\aa}}\ \emph {et~al.}(2007)\citenamefont
  {S{\o}nden{\aa}}, \citenamefont {St{\o}len}, \citenamefont {Ravindran},\ and\
  \citenamefont {Grande}}]{sondenaa2007heat}%
  \BibitemOpen
  \bibfield  {author} {\bibinfo {author} {\bibfnamefont {R.}~\bibnamefont
  {S{\o}nden{\aa}}}, \bibinfo {author} {\bibfnamefont {S.}~\bibnamefont
  {St{\o}len}}, \bibinfo {author} {\bibfnamefont {P.}~\bibnamefont
  {Ravindran}},\ and\ \bibinfo {author} {\bibfnamefont {T.}~\bibnamefont
  {Grande}},\ }\bibfield  {title} {\bibinfo {title} {Heat capacity and lattice
  dynamics of cubic and hexagonal sr mn o 3: Calorimetry and density functional
  theory simulations},\ }\href@noop {} {\bibfield  {journal} {\bibinfo
  {journal} {Physical Review B—Condensed Matter and Materials Physics}\
  }\textbf {\bibinfo {volume} {75}},\ \bibinfo {pages} {214307} (\bibinfo
  {year} {2007})}\BibitemShut {NoStop}%
\bibitem [{\citenamefont {Lin}\ \emph {et~al.}(2000)\citenamefont {Lin},
  \citenamefont {Chun}, \citenamefont {Salamon}, \citenamefont {Tomioka},\ and\
  \citenamefont {Tokura}}]{lin2000magnetic}%
  \BibitemOpen
  \bibfield  {author} {\bibinfo {author} {\bibfnamefont {P.}~\bibnamefont
  {Lin}}, \bibinfo {author} {\bibfnamefont {S.}~\bibnamefont {Chun}}, \bibinfo
  {author} {\bibfnamefont {M.}~\bibnamefont {Salamon}}, \bibinfo {author}
  {\bibfnamefont {Y.}~\bibnamefont {Tomioka}},\ and\ \bibinfo {author}
  {\bibfnamefont {Y.}~\bibnamefont {Tokura}},\ }\bibfield  {title} {\bibinfo
  {title} {Magnetic heat capacity in lanthanum manganite single crystals},\
  }\href@noop {} {\bibfield  {journal} {\bibinfo  {journal} {Journal of Applied
  Physics}\ }\textbf {\bibinfo {volume} {87}},\ \bibinfo {pages} {5825}
  (\bibinfo {year} {2000})}\BibitemShut {NoStop}%
\bibitem [{\citenamefont {Tishin}\ \emph {et~al.}(1999)\citenamefont {Tishin},
  \citenamefont {Gschneidner~Jr},\ and\ \citenamefont
  {Pecharsky}}]{tishin1999magnetocaloric}%
  \BibitemOpen
  \bibfield  {author} {\bibinfo {author} {\bibfnamefont {A.}~\bibnamefont
  {Tishin}}, \bibinfo {author} {\bibfnamefont {K.}~\bibnamefont
  {Gschneidner~Jr}},\ and\ \bibinfo {author} {\bibfnamefont {V.}~\bibnamefont
  {Pecharsky}},\ }\bibfield  {title} {\bibinfo {title} {Magnetocaloric effect
  and heat capacity in the phase-transition region},\ }\href@noop {} {\bibfield
   {journal} {\bibinfo  {journal} {Physical Review B}\ }\textbf {\bibinfo
  {volume} {59}},\ \bibinfo {pages} {503} (\bibinfo {year} {1999})}\BibitemShut
  {NoStop}%
\bibitem [{\citenamefont {Walsh}\ \emph {et~al.}(2022)\citenamefont {Walsh},
  \citenamefont {Asta},\ and\ \citenamefont {Wang}}]{Walsh2022}%
  \BibitemOpen
  \bibfield  {author} {\bibinfo {author} {\bibfnamefont {F.}~\bibnamefont
  {Walsh}}, \bibinfo {author} {\bibfnamefont {M.}~\bibnamefont {Asta}},\ and\
  \bibinfo {author} {\bibfnamefont {L.-W.}\ \bibnamefont {Wang}},\ }\bibfield
  {title} {\bibinfo {title} {{Realistic magnetic thermodynamics by local
  quantization of a semiclassical Heisenberg model}},\ }\href
  {https://doi.org/10.1038/s41524-022-00875-8} {\bibfield  {journal} {\bibinfo
  {journal} {npj Computational Materials}\ }\textbf {\bibinfo {volume} {8}},\
  \bibinfo {pages} {186} (\bibinfo {year} {2022})}\BibitemShut {NoStop}%
\bibitem [{\citenamefont {Mendive-Tapia}\ \emph {et~al.}(2022)\citenamefont
  {Mendive-Tapia}, \citenamefont {Neugebauer},\ and\ \citenamefont
  {Hickel}}]{Mendive-Tapia2022}%
  \BibitemOpen
  \bibfield  {author} {\bibinfo {author} {\bibfnamefont {E.}~\bibnamefont
  {Mendive-Tapia}}, \bibinfo {author} {\bibfnamefont {J.}~\bibnamefont
  {Neugebauer}},\ and\ \bibinfo {author} {\bibfnamefont {T.}~\bibnamefont
  {Hickel}},\ }\bibfield  {title} {\bibinfo {title} {{Ab initio calculation of
  the magnetic Gibbs free energy of materials using magnetically constrained
  supercells}},\ }\href {https://doi.org/10.1103/PhysRevB.105.064425}
  {\bibfield  {journal} {\bibinfo  {journal} {Physical Review B}\ }\textbf
  {\bibinfo {volume} {105}},\ \bibinfo {pages} {064425} (\bibinfo {year}
  {2022})},\ \Eprint {https://arxiv.org/abs/2202.11492} {arXiv:2202.11492}
  \BibitemShut {NoStop}%
\bibitem [{\citenamefont {Sangeetha}\ \emph {et~al.}(2016)\citenamefont
  {Sangeetha}, \citenamefont {Cuervo-Reyes}, \citenamefont {Pandey},\ and\
  \citenamefont {Johnston}}]{Sangeetha2016}%
  \BibitemOpen
  \bibfield  {author} {\bibinfo {author} {\bibfnamefont {N.~S.}\ \bibnamefont
  {Sangeetha}}, \bibinfo {author} {\bibfnamefont {E.}~\bibnamefont
  {Cuervo-Reyes}}, \bibinfo {author} {\bibfnamefont {A.}~\bibnamefont
  {Pandey}},\ and\ \bibinfo {author} {\bibfnamefont {D.~C.}\ \bibnamefont
  {Johnston}},\ }\bibfield  {title} {\bibinfo {title} {{EuCo2P2: A Model
  Molecular-Field Helical Heisenberg Antiferromagnet}},\ }\href
  {https://doi.org/10.1103/PhysRevB.94.014422} {\bibfield  {journal} {\bibinfo
  {journal} {Physical Review B}\ }\textbf {\bibinfo {volume} {94}},\ \bibinfo
  {pages} {014422} (\bibinfo {year} {2016})},\ \Eprint
  {https://arxiv.org/abs/1607.00247} {arXiv:1607.00247} \BibitemShut {NoStop}%
\bibitem [{\citenamefont {Horton}\ \emph {et~al.}(2019)\citenamefont {Horton},
  \citenamefont {Montoya}, \citenamefont {Liu},\ and\ \citenamefont
  {Persson}}]{horton2019high}%
  \BibitemOpen
  \bibfield  {author} {\bibinfo {author} {\bibfnamefont {M.~K.}\ \bibnamefont
  {Horton}}, \bibinfo {author} {\bibfnamefont {J.~H.}\ \bibnamefont {Montoya}},
  \bibinfo {author} {\bibfnamefont {M.}~\bibnamefont {Liu}},\ and\ \bibinfo
  {author} {\bibfnamefont {K.~A.}\ \bibnamefont {Persson}},\ }\bibfield
  {title} {\bibinfo {title} {High-throughput prediction of the ground-state
  collinear magnetic order of inorganic materials using density functional
  theory},\ }\href@noop {} {\bibfield  {journal} {\bibinfo  {journal} {npj
  Computational Materials}\ }\textbf {\bibinfo {volume} {5}},\ \bibinfo {pages}
  {64} (\bibinfo {year} {2019})}\BibitemShut {NoStop}%
\bibitem [{\citenamefont {Gyorffy}\ \emph {et~al.}(1985)\citenamefont
  {Gyorffy}, \citenamefont {Pindor}, \citenamefont {Staunton}, \citenamefont
  {Stocks},\ and\ \citenamefont {Winter}}]{gyorffy1985first}%
  \BibitemOpen
  \bibfield  {author} {\bibinfo {author} {\bibfnamefont {B.}~\bibnamefont
  {Gyorffy}}, \bibinfo {author} {\bibfnamefont {A.}~\bibnamefont {Pindor}},
  \bibinfo {author} {\bibfnamefont {J.}~\bibnamefont {Staunton}}, \bibinfo
  {author} {\bibfnamefont {G.}~\bibnamefont {Stocks}},\ and\ \bibinfo {author}
  {\bibfnamefont {H.}~\bibnamefont {Winter}},\ }\bibfield  {title} {\bibinfo
  {title} {A first-principles theory of ferromagnetic phase transitions in
  metals},\ }\href@noop {} {\bibfield  {journal} {\bibinfo  {journal} {Journal
  of Physics F: Metal Physics}\ }\textbf {\bibinfo {volume} {15}},\ \bibinfo
  {pages} {1337} (\bibinfo {year} {1985})}\BibitemShut {NoStop}%
\bibitem [{\citenamefont {Anderson}(1950)}]{anderson1950antiferromagnetism}%
  \BibitemOpen
  \bibfield  {author} {\bibinfo {author} {\bibfnamefont {P.~W.}\ \bibnamefont
  {Anderson}},\ }\bibfield  {title} {\bibinfo {title} {Antiferromagnetism.
  theory of superexchange interaction},\ }\href@noop {} {\bibfield  {journal}
  {\bibinfo  {journal} {Physical Review}\ }\textbf {\bibinfo {volume} {79}},\
  \bibinfo {pages} {350} (\bibinfo {year} {1950})}\BibitemShut {NoStop}%
\bibitem [{\citenamefont {Lines}(1964)}]{lines1964green}%
  \BibitemOpen
  \bibfield  {author} {\bibinfo {author} {\bibfnamefont {M.}~\bibnamefont
  {Lines}},\ }\bibfield  {title} {\bibinfo {title} {Green functions in the
  theory of antiferromagnetism},\ }\href@noop {} {\bibfield  {journal}
  {\bibinfo  {journal} {Physical Review}\ }\textbf {\bibinfo {volume} {135}},\
  \bibinfo {pages} {A1336} (\bibinfo {year} {1964})}\BibitemShut {NoStop}%
\bibitem [{\citenamefont {Korotin}\ \emph {et~al.}(2015)\citenamefont
  {Korotin}, \citenamefont {Mazurenko}, \citenamefont {Anisimov},\ and\
  \citenamefont {Streltsov}}]{korotin2015calculation}%
  \BibitemOpen
  \bibfield  {author} {\bibinfo {author} {\bibfnamefont {D.~M.}\ \bibnamefont
  {Korotin}}, \bibinfo {author} {\bibfnamefont {V.}~\bibnamefont {Mazurenko}},
  \bibinfo {author} {\bibfnamefont {V.}~\bibnamefont {Anisimov}},\ and\
  \bibinfo {author} {\bibfnamefont {S.}~\bibnamefont {Streltsov}},\ }\bibfield
  {title} {\bibinfo {title} {Calculation of exchange constants of the
  heisenberg model in plane-wave-based methods using the green's function
  approach},\ }\href@noop {} {\bibfield  {journal} {\bibinfo  {journal}
  {Physical Review B}\ }\textbf {\bibinfo {volume} {91}},\ \bibinfo {pages}
  {224405} (\bibinfo {year} {2015})}\BibitemShut {NoStop}%
\bibitem [{\citenamefont {Mare}\ \emph {et~al.}(2017)\citenamefont {Mare},
  \citenamefont {Moreira},\ and\ \citenamefont
  {Rossi}}]{mare2017nonstationary}%
  \BibitemOpen
  \bibfield  {author} {\bibinfo {author} {\bibfnamefont {D.~S.}\ \bibnamefont
  {Mare}}, \bibinfo {author} {\bibfnamefont {F.}~\bibnamefont {Moreira}},\ and\
  \bibinfo {author} {\bibfnamefont {R.}~\bibnamefont {Rossi}},\ }\bibfield
  {title} {\bibinfo {title} {Nonstationary z-score measures},\ }\href@noop {}
  {\bibfield  {journal} {\bibinfo  {journal} {European Journal of Operational
  Research}\ }\textbf {\bibinfo {volume} {260}},\ \bibinfo {pages} {348}
  (\bibinfo {year} {2017})}\BibitemShut {NoStop}%
\bibitem [{\citenamefont {Ahsanullah}\ \emph {et~al.}(2014)\citenamefont
  {Ahsanullah}, \citenamefont {Kibria},\ and\ \citenamefont
  {Shakil}}]{ahsanullah2014normal}%
  \BibitemOpen
  \bibfield  {author} {\bibinfo {author} {\bibfnamefont {M.}~\bibnamefont
  {Ahsanullah}}, \bibinfo {author} {\bibfnamefont {B.~G.}\ \bibnamefont
  {Kibria}},\ and\ \bibinfo {author} {\bibfnamefont {M.}~\bibnamefont
  {Shakil}},\ }\href@noop {} {\emph {\bibinfo {title} {Normal and student's t
  distributions and their applications}}},\ Vol.~\bibinfo {volume} {4}\
  (\bibinfo  {publisher} {Springer},\ \bibinfo {year} {2014})\BibitemShut
  {NoStop}%
\bibitem [{\citenamefont {Martin}\ \emph {et~al.}(2018)\citenamefont {Martin},
  \citenamefont {Bieder}, \citenamefont {Prokhorenko},\ and\ \citenamefont
  {Ghosez}}]{martin2018multibinit}%
  \BibitemOpen
  \bibfield  {author} {\bibinfo {author} {\bibfnamefont {A.}~\bibnamefont
  {Martin}}, \bibinfo {author} {\bibfnamefont {J.}~\bibnamefont {Bieder}},
  \bibinfo {author} {\bibfnamefont {S.}~\bibnamefont {Prokhorenko}},\ and\
  \bibinfo {author} {\bibfnamefont {P.}~\bibnamefont {Ghosez}},\ }\bibfield
  {title} {\bibinfo {title} {The multibinit software project},\ }in\ \href@noop
  {} {\emph {\bibinfo {booktitle} {APS March Meeting Abstracts}}},\ Vol.\
  \bibinfo {volume} {2018}\ (\bibinfo {year} {2018})\ pp.\ \bibinfo {pages}
  {K02--004}\BibitemShut {NoStop}%
\bibitem [{\citenamefont {Soler}\ \emph {et~al.}(2002)\citenamefont {Soler},
  \citenamefont {Artacho}, \citenamefont {Gale}, \citenamefont {Garc{\'\i}a},
  \citenamefont {Junquera}, \citenamefont {Ordej{\'o}n},\ and\ \citenamefont
  {S{\'a}nchez-Portal}}]{soler2002siesta}%
  \BibitemOpen
  \bibfield  {author} {\bibinfo {author} {\bibfnamefont {J.~M.}\ \bibnamefont
  {Soler}}, \bibinfo {author} {\bibfnamefont {E.}~\bibnamefont {Artacho}},
  \bibinfo {author} {\bibfnamefont {J.~D.}\ \bibnamefont {Gale}}, \bibinfo
  {author} {\bibfnamefont {A.}~\bibnamefont {Garc{\'\i}a}}, \bibinfo {author}
  {\bibfnamefont {J.}~\bibnamefont {Junquera}}, \bibinfo {author}
  {\bibfnamefont {P.}~\bibnamefont {Ordej{\'o}n}},\ and\ \bibinfo {author}
  {\bibfnamefont {D.}~\bibnamefont {S{\'a}nchez-Portal}},\ }\bibfield  {title}
  {\bibinfo {title} {The siesta method for ab initio order-nmaterials
  simulation},\ }\href@noop {} {\bibfield  {journal} {\bibinfo  {journal}
  {Journal of Physics: Condensed Matter}\ }\textbf {\bibinfo {volume} {14}},\
  \bibinfo {pages} {2745} (\bibinfo {year} {2002})}\BibitemShut {NoStop}%
\bibitem [{\citenamefont {García}\ \emph {et~al.}(2018)\citenamefont
  {García}, \citenamefont {Verstraete}, \citenamefont {Pouillon},\ and\
  \citenamefont {Junquera}}]{Garc_a_2018}%
  \BibitemOpen
  \bibfield  {author} {\bibinfo {author} {\bibfnamefont {A.}~\bibnamefont
  {García}}, \bibinfo {author} {\bibfnamefont {M.~J.}\ \bibnamefont
  {Verstraete}}, \bibinfo {author} {\bibfnamefont {Y.}~\bibnamefont
  {Pouillon}},\ and\ \bibinfo {author} {\bibfnamefont {J.}~\bibnamefont
  {Junquera}},\ }\bibfield  {title} {\bibinfo {title} {The psml format and
  library for norm-conserving pseudopotential data curation and
  interoperability},\ }\href {https://doi.org/10.1016/j.cpc.2018.02.011}
  {\bibfield  {journal} {\bibinfo  {journal} {Computer Physics Communications}\
  }\textbf {\bibinfo {volume} {227}},\ \bibinfo {pages} {51–71} (\bibinfo
  {year} {2018})}\BibitemShut {NoStop}%
\bibitem [{\citenamefont {Perdew}\ \emph {et~al.}(1996)\citenamefont {Perdew},
  \citenamefont {Burke},\ and\ \citenamefont
  {Ernzerhof}}]{perdew1996generalized}%
  \BibitemOpen
  \bibfield  {author} {\bibinfo {author} {\bibfnamefont {J.~P.}\ \bibnamefont
  {Perdew}}, \bibinfo {author} {\bibfnamefont {K.}~\bibnamefont {Burke}},\ and\
  \bibinfo {author} {\bibfnamefont {M.}~\bibnamefont {Ernzerhof}},\ }\bibfield
  {title} {\bibinfo {title} {Generalized gradient approximation made simple},\
  }\href@noop {} {\bibfield  {journal} {\bibinfo  {journal} {Physical review
  letters}\ }\textbf {\bibinfo {volume} {77}},\ \bibinfo {pages} {3865}
  (\bibinfo {year} {1996})}\BibitemShut {NoStop}%
\bibitem [{\citenamefont {He}\ \emph {et~al.}(2021)\citenamefont {He},
  \citenamefont {Helbig}, \citenamefont {Verstraete},\ and\ \citenamefont
  {Bousquet}}]{he2021tb2j}%
  \BibitemOpen
  \bibfield  {author} {\bibinfo {author} {\bibfnamefont {X.}~\bibnamefont
  {He}}, \bibinfo {author} {\bibfnamefont {N.}~\bibnamefont {Helbig}}, \bibinfo
  {author} {\bibfnamefont {M.~J.}\ \bibnamefont {Verstraete}},\ and\ \bibinfo
  {author} {\bibfnamefont {E.}~\bibnamefont {Bousquet}},\ }\bibfield  {title}
  {\bibinfo {title} {Tb2j: A python package for computing magnetic interaction
  parameters},\ }\href@noop {} {\bibfield  {journal} {\bibinfo  {journal}
  {Computer Physics Communications}\ }\textbf {\bibinfo {volume} {264}},\
  \bibinfo {pages} {107938} (\bibinfo {year} {2021})}\BibitemShut {NoStop}%
\bibitem [{\citenamefont {Lakshmanan}(2011)}]{lakshmanan2011fascinating}%
  \BibitemOpen
  \bibfield  {author} {\bibinfo {author} {\bibfnamefont {M.}~\bibnamefont
  {Lakshmanan}},\ }\bibfield  {title} {\bibinfo {title} {The fascinating world
  of the landau--lifshitz--gilbert equation: an overview},\ }\href@noop {}
  {\bibfield  {journal} {\bibinfo  {journal} {Philosophical Transactions of the
  Royal Society A: Mathematical, Physical and Engineering Sciences}\ }\textbf
  {\bibinfo {volume} {369}},\ \bibinfo {pages} {1280} (\bibinfo {year}
  {2011})}\BibitemShut {NoStop}%
\bibitem [{\citenamefont {Hagl{\"o}f}\ \emph {et~al.}(2021)\citenamefont
  {Hagl{\"o}f}, \citenamefont {Blomqvist}, \citenamefont {Ruban},\ and\
  \citenamefont {Selleby}}]{haglof2021calphad}%
  \BibitemOpen
  \bibfield  {author} {\bibinfo {author} {\bibfnamefont {F.}~\bibnamefont
  {Hagl{\"o}f}}, \bibinfo {author} {\bibfnamefont {A.}~\bibnamefont
  {Blomqvist}}, \bibinfo {author} {\bibfnamefont {A.}~\bibnamefont {Ruban}},\
  and\ \bibinfo {author} {\bibfnamefont {M.}~\bibnamefont {Selleby}},\
  }\bibfield  {title} {\bibinfo {title} {Calphad: Method for calculation of
  finite temperature thermodynamic properties for magnetic allotropes—case
  study on fe, co and ni},\ }\href@noop {} {\bibfield  {journal} {\bibinfo
  {journal} {Calphad}\ }\textbf {\bibinfo {volume} {74}},\ \bibinfo {pages}
  {102320} (\bibinfo {year} {2021})}\BibitemShut {NoStop}%
\bibitem [{\citenamefont {He}\ \emph {et~al.}(2022)\citenamefont {He},
  \citenamefont {Hagl{\"o}f}, \citenamefont {Chen}, \citenamefont {Blomqvist},\
  and\ \citenamefont {Selleby}}]{he2022third}%
  \BibitemOpen
  \bibfield  {author} {\bibinfo {author} {\bibfnamefont {Z.}~\bibnamefont
  {He}}, \bibinfo {author} {\bibfnamefont {F.}~\bibnamefont {Hagl{\"o}f}},
  \bibinfo {author} {\bibfnamefont {Q.}~\bibnamefont {Chen}}, \bibinfo {author}
  {\bibfnamefont {A.}~\bibnamefont {Blomqvist}},\ and\ \bibinfo {author}
  {\bibfnamefont {M.}~\bibnamefont {Selleby}},\ }\bibfield  {title} {\bibinfo
  {title} {A third generation calphad description of fe: Revisions of fcc, hcp
  and liquid},\ }\href@noop {} {\bibfield  {journal} {\bibinfo  {journal}
  {Journal of Phase Equilibria and Diffusion}\ }\textbf {\bibinfo {volume}
  {43}},\ \bibinfo {pages} {287} (\bibinfo {year} {2022})}\BibitemShut
  {NoStop}%
\bibitem [{\citenamefont {Li}\ \emph {et~al.}(2021)\citenamefont {Li},
  \citenamefont {Fu},\ and\ \citenamefont {Schneider}}]{li2021effects}%
  \BibitemOpen
  \bibfield  {author} {\bibinfo {author} {\bibfnamefont {K.}~\bibnamefont
  {Li}}, \bibinfo {author} {\bibfnamefont {C.-C.}\ \bibnamefont {Fu}},\ and\
  \bibinfo {author} {\bibfnamefont {A.}~\bibnamefont {Schneider}},\ }\bibfield
  {title} {\bibinfo {title} {Effects of magnetic excitations and transitions on
  vacancy formation: Cases of fcc fe and ni compared to bcc fe},\ }\href@noop
  {} {\bibfield  {journal} {\bibinfo  {journal} {Physical Review B}\ }\textbf
  {\bibinfo {volume} {104}},\ \bibinfo {pages} {104406} (\bibinfo {year}
  {2021})}\BibitemShut {NoStop}%
\bibitem [{\citenamefont {Mostovoy}(2024)}]{mostovoy2024multiferroics}%
  \BibitemOpen
  \bibfield  {author} {\bibinfo {author} {\bibfnamefont {M.}~\bibnamefont
  {Mostovoy}},\ }\bibfield  {title} {\bibinfo {title} {Multiferroics: Different
  routes to magnetoelectric coupling},\ }\href@noop {} {\bibfield  {journal}
  {\bibinfo  {journal} {npj Spintronics}\ }\textbf {\bibinfo {volume} {2}},\
  \bibinfo {pages} {18} (\bibinfo {year} {2024})}\BibitemShut {NoStop}%
\bibitem [{\citenamefont {Daalderop}\ \emph {et~al.}(1991)\citenamefont
  {Daalderop}, \citenamefont {Kelly},\ and\ \citenamefont
  {Schuurmans}}]{daalderop1991magnetocrystalline}%
  \BibitemOpen
  \bibfield  {author} {\bibinfo {author} {\bibfnamefont {G.}~\bibnamefont
  {Daalderop}}, \bibinfo {author} {\bibfnamefont {P.}~\bibnamefont {Kelly}},\
  and\ \bibinfo {author} {\bibfnamefont {M.}~\bibnamefont {Schuurmans}},\
  }\bibfield  {title} {\bibinfo {title} {Magnetocrystalline anisotropy and
  orbital moments in transition-metal compounds},\ }\href@noop {} {\bibfield
  {journal} {\bibinfo  {journal} {Physical Review B}\ }\textbf {\bibinfo
  {volume} {44}},\ \bibinfo {pages} {12054} (\bibinfo {year}
  {1991})}\BibitemShut {NoStop}%
\bibitem [{\citenamefont {Inzani}\ \emph {et~al.}(2022)\citenamefont {Inzani},
  \citenamefont {Pokhrel}, \citenamefont {Leclerc}, \citenamefont {Clemens},
  \citenamefont {Ramkumar}, \citenamefont {Griffin},\ and\ \citenamefont
  {Nowadnick}}]{inzani2022manipulation}%
  \BibitemOpen
  \bibfield  {author} {\bibinfo {author} {\bibfnamefont {K.}~\bibnamefont
  {Inzani}}, \bibinfo {author} {\bibfnamefont {N.}~\bibnamefont {Pokhrel}},
  \bibinfo {author} {\bibfnamefont {N.}~\bibnamefont {Leclerc}}, \bibinfo
  {author} {\bibfnamefont {Z.}~\bibnamefont {Clemens}}, \bibinfo {author}
  {\bibfnamefont {S.~P.}\ \bibnamefont {Ramkumar}}, \bibinfo {author}
  {\bibfnamefont {S.~M.}\ \bibnamefont {Griffin}},\ and\ \bibinfo {author}
  {\bibfnamefont {E.~A.}\ \bibnamefont {Nowadnick}},\ }\bibfield  {title}
  {\bibinfo {title} {Manipulation of spin orientation via ferroelectric
  switching in fe-doped bi 2 wo 6 from first principles},\ }\href@noop {}
  {\bibfield  {journal} {\bibinfo  {journal} {Physical Review B}\ }\textbf
  {\bibinfo {volume} {105}},\ \bibinfo {pages} {054434} (\bibinfo {year}
  {2022})}\BibitemShut {NoStop}%
\end{thebibliography}%
